\documentclass[preprint2]{aastex7} 
\usepackage{enumitem}
\usepackage{verbatim}
\usepackage{amsmath}

\shorttitle{FRBs probe Galaxy Evolution}
\shortauthors{Ll. Mas-Ribas} 

\begin{document}

\title{Fast Radio Bursts probe Galaxy Evolution:\\ 
Evidence and implications of a redshift-dependent FRB host DM} 

\author[orcid=0000-0003-4584-8841]{Lluis Mas-Ribas}
\affiliation{Department of Astronomy and Astrophysics, University of California, \\1156 High Street, Santa Cruz, CA 95064, USA}

\email[show]{lluismasribas@gmail.com}

\begin{abstract}

The redshift evolution of ionized gas  in the full
galaxy-halo system is a central open question in galaxy formation, because no 
existing observable is simultaneously sensitive to all ionized phases.
Here we explore fast radio bursts (FRBs) 
as a  probe of the density evolution of this gas  
through the redshift dependence of the 
FRB host dispersion measure, ${\rm DM_{host}}(z)  \propto (1+z)^{n_z}$. 
The host DM denotes the  electron 
column density of all ionized gas in the host along the FRB sightline, providing 
a  unified tracer that complements existing 
phase-specific  diagnostics. 
We apply a forward-modeling framework that accounts for instrumental 
effects  to 90 localized FRBs (69 with confirmed 
host redshifts) from the DSA and ASKAP/CRAFT ICS surveys. Our  
inference yields $n_z = 1.62^{+1.48}_{-1.57}$, ruling out the 
non-evolving scenario ($n_z = 0$) at $1\,\sigma$, with both  
datasets independently favoring $n_z > 0$. The main $n_z$ degeneracy  
is with the mean host DM and parameters such as $H_0$, highlighting the 
need to account for a host evolution in inference analyses and DM-based host redshift estimates; overestimating  redshifts by up to $\Delta z \approx 0.3$ for
DM$_{\rm EG} \sim 1000 - 2000\,{\rm pc\,cm^{-3}}$ otherwise. About 100 ($300-350$) localized hosts from MeerTRAP in coherent mode (DSA/CRAFT) will yield $n_z$
uncertainties of $\sim0.7$. Precise $n_z$ measurements  compared 
with the  evolution of individual phases and galaxy 
scaling relations will shed light on ionized gas evolution  in
galaxies and halos, informing the 
dominant phase, the driver of the overall evolution, and FRB progenitor channels.

\end{abstract}



\section{Introduction}\label{sec:intro}

The ionized gas in galaxies and   halos encodes a 
detailed record of the physical processes that drive galaxy evolution. 
The gas density, in particular, traces the ionization state, 
temperature, and mass budget of the gas \citep{Osterbrock2006}, and 
its evolution  with cosmic time reflects the cumulative action 
of star formation,  gravitational accretion, and stellar and active galactic 
nuclei (AGN) feedback \citep[e.g.,][]{Somerville2015}. 
Measuring the density of ionized gas  across 
a wide range of redshifts, environments, and ionized phases is therefore 
a powerful avenue for understanding how galaxies assemble and evolve. 
However, no single observable currently integrates all phases simultaneously,
making it difficult to build a unified picture of ionized gas
evolution. 
In this paper, we explore fast radio 
bursts (FRBs) as a novel and complementary probe of the density evolution 
of \textit{all} 
ionized gas in the galaxy-halo system.

\subsection{Density evolution in galaxies and halos}

In galaxies, diffuse ionized gas (DIG)\footnote{In this paper 
we refer to galaxies other than the Milky Way. In our Galaxy, the diffuse 
ionized component is often named warm ionized medium (WIM). Furthermore, the 
distribution of ionized gas in the Milky Way is modeled in detail from 
observations of pulsars \citep{Cordes2002,Cordes2003,Ocker2026}.} with temperature 
$T\sim10^4$\,K, electron density $n_e\sim 0.1-1\,{\rm cm^{-3}}$, and 
extending out to $\approx 1$\,kpc above the plane of the interstellar medium 
(ISM), occupies 
$\approx 90\%$ of the volume. Star formation activity is responsible for the 
extent of the DIG and for its $n_e$ redshift evolution, yielding 
$n_e \propto (1+z)^{\approx 2.5}$ up to the peak epoch of star formation 
at $z\sim 2-3$ 
\citep{Haffner2009,Levy2019,McClymont2024}. 

Embedded within the DIG, \ion{H}{2} regions contain denser 
($n_e\gtrsim10-100~{\rm cm^{-3}}$) ionized gas. 
Its $n_e$ evolution is steep between $z=0$ and 
$z\sim2-3$, with $n_e \propto (1+z)^{\sim 2}$, while a flatter trend 
with $n_e \propto (1+z)^{1-1.5}$ is favored at higher redshifts up to 
$z\sim10-11$ \citep{Abdurrouf2024,Topping2025,Martinez2025}. 
Despite general consensus on these values, the driver of this \ion{H}{2} 
density evolution remains debated: \cite{Isobe2023} found that the increase 
of $n_e$ with redshift is unrelated to the corresponding increase of galaxy 
stellar mass or star formation. Contrarily, \cite{Kaasinen2017} argued that 
the evolution of the cosmic star formation rate is the reason for the 
observed trend, and \cite{Topping2025} found a weak $n_e$ correlation with 
star formation rate surface density but not with stellar mass 
\citep[see also][]{Shirazi2014,Shimakawa2015,Davies2021,Martinez2025,Li2025}. 
Several studies further suggest that the increase of galaxy density with 
redshift may be, at least partially, driven by the corresponding decrease in 
galaxy size \citep[e.g.,][]{Shibuya2015}, shown to follow 
$r \propto (1+z)^p$ with $p\sim -0.8$ and $p\sim-0.4$ for star-forming 
and quenched galaxies, respectively \citep{Ormerod2024}.

In halos, the ionized gas extending out to the virial radius is referred 
to as the circumgalactic medium (CGM). It comprises cool and warm phases at 
temperatures $T\sim10^4-10^6$\,K and densities 
$n_e\sim10^{-2}-10^{-4}~{\rm cm^{-3}}$, as well as a hot phase at 
$T\gtrsim10^6$\,K and $n_e \lesssim 10^{-4}-10^{-5}~{\rm cm^{-3}}$ 
\citep{Tumlinson2017,Fumagalli2024}. This gas is driven into halos via 
gravitational accretion, and by stellar and AGN feedback from the central 
galaxy. The overall halo gas density traces the cosmic expansion as 
$n_e \propto (1+z)^3$, although significant variation may exist with galaxy 
mass, type, and star-formation and feedback activity 
\citep{Birnboim2003,Vandevoort2012,Correa2018,Sorini2024}. The CGM size, 
in turn, evolves with the halo virial radius as $r\propto (1+z)^{-1}$.

\begin{figure*}\centering
\includegraphics[width=0.95\textwidth]{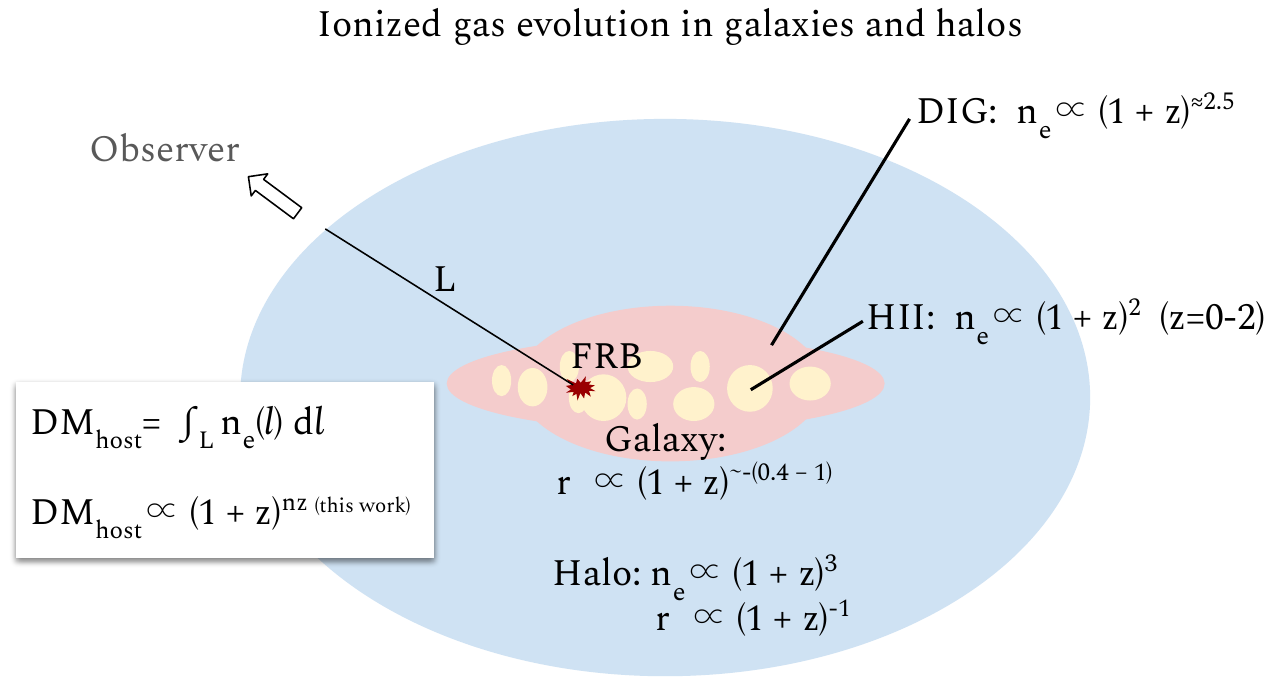}
\caption{Redshift evolution of ionized gas density and size 
(radius) of galaxies (constituted by the diffuse ionized gas, DIG, 
and \ion{H}{2} regions, for redshifts $z\sim 0-2$) and halos. 
The diversity of trends and the ongoing debate about which
physical processes drive them motivates the use of a single
observable  that integrates all ionized
phases simultaneously. Here 
we investigate the redshift evolution of ${\rm DM_{host}}$ 
as a complementary probe of galaxy evolution, where ${\rm DM_{host}}$  
denotes the column density of all ionized gas along the FRB sightline through 
the host environment.  Not to scale.}
\label{fig:cart}
\end{figure*}

Figure~\ref{fig:cart} illustrates the density and size evolutions of the 
aforementioned environments, together with the information provided by FRBs 
discussed in the next section. The diversity of trends across the DIG, 
\ion{H}{2} regions, and CGM -- and the ongoing debate about the processes 
driving them -- motivates the use of a unified, global tracer that integrates 
the total ionized gas budget of the galaxy-halo system across cosmic time. 
Although such a probe does not resolve the contribution of the individual 
components, it may reveal the dominant phase, the driver of the overall 
density evolution and, given the close 
interrelation between the galaxy and its halo \citep{Lilly2013,Walter2020}, it 
serves as a natural complement to existing studies that 
typically focus on one medium at a time.

\subsection{FRBs as probes of galaxy evolution}

FRBs are luminous $\sim$millisecond-duration extragalactic radio  
pulses of yet unknown origin, currently detected up to $z=2.15$ 
\citep{Caleb2026}, with a fraction of $\sim2-3\%$ showing repetition 
\citep{Kharel2026,Cook2026}. The currently identified  FRB hosts 
span a wide variety of galaxy types, from dwarf to massive structures and 
from quenched to 
star-forming, with a current preference for massive star-forming systems 
\citep{Niu2022,Gordon2023,Hewitt2024,Eftekari2024,Sharma2024,Amiri2025,Gordon2025,bright2025,pastor2026}. 

FRBs are particularly well suited to probe all phases of the ionized gas  
since they encode the \textit{total} electron column density along their 
sightline in the observable referred to as dispersion measure (DM), which  
is precisely measured from the frequency-dependent delay in the arrival time 
of the radio signal \citep{Petroff2019}. This stands in contrast to conventional probes of 
ionized gas: emission-line diagnostics of galaxies are primarily sensitive to 
the dense \ion{H}{2} phase and require rest-frame optical and/or ultraviolet 
coverage,  as well as assumptions on pressure or temperature for carrying  photoionization modeling   \citep{Wang2004,Kewley2019,Xing2026}; absorption-line 
spectroscopy of background quasars probes pencil-beam sightlines through the 
cool and warm CGM but is limited by the availability of bright background 
sources, and observations of halo gas in emission, although promising, are 
still in the early stages \citep{Fumagalli2024}; finally,  
X-ray or Sunyaev-Zel'dovich observations access the hot CGM but are not 
sensitive to the full range of spatial scales, masses, and redshifts 
\citep[][and references therein]{Masribas2017}. FRBs, by contrast, provide a 
single precisely measured observable that is simultaneously sensitive to all 
ionized phases and equally applicable from low to high redshift \citep[see, e.g.,][for FRBs probing intervening foreground halos]{Prochaska2019,Shin2024,Lanman2025,Kahinga2026}. 
Furthermore, the growing sample of localized FRBs with confirmed host redshifts 
-- currently $\sim120$, with rapid growth expected from the ongoing CHIME/FRB Outriggers 
\citep{Amiriout2025}, as well as upcoming CHORD \citep{chord2019} and DSA-2000 \citep{dsa2000}, 
together with large spectroscopic surveys such as DESI \citep{Desi2026} and 
EUCLID \citep{Euclid2025} -- is making FRBs a statistically viable tool 
for this class of analysis.

The rest-frame DM of a host with FRB path length $L$ (Figure~\ref{fig:cart}) 
equates   
\begin{equation}
    {\rm DM_{host}} = \int_L n_e(l)\, {\rm d}l~. 
\end{equation}
The distribution of ${\rm DM_{host}}$ values 
is commonly modeled as a log-normal  with 
mean value $\sim 100\,{\rm pc\,cm^{-3}}$ and with no redshift evolution 
assumed in the majority of analyses \citep[although see, e.g.,][]{Wu2022,Zhang2024,Mo2025,Reischke2025}. 
In this work, we allow for a redshift-dependent host DM of the form
\begin{equation}\label{eq:dmz}
   {\rm DM_{host}}(z) = {\rm DM_{host,0}} \,(1+z)^{n_z}~,
\end{equation}
where $n_z$ parameterizes the redshift evolution and $n_z=0$ 
corresponds to the non-evolving case. The term ${\rm DM_{host,0}}$ 
denotes the aforementioned log-normal distribution at $z=0$. 

Explorations of a host DM redshift evolution have so far relied mostly 
on simulations. Works using Illustris and IllustrisTNG \citep{Nelson2018} 
by \cite{Jaroszynski2020,Zhang2020,Mo2023,Kovacs2024} generally find an 
increase of host DM with redshift, with $n_z \sim 0.6-1.8$ between 
$z\sim 0-2$ \citep[although see][for a decreasing trend]{Theis2024}. The 
exact values depend on simulation setup, galaxy stellar mass, location 
of the FRB within the host, and galaxy type (i.e., star-forming galaxies 
showing steeper evolutions than quenched ones).  
Interestingly, \cite{Orr2024} used high-resolution FIRE-2 simulations 
\citep{Hopkins2015,wetzel2023} and found that the dominant driver of 
the host 
DM evolution is not stellar mass or star formation but  the host galaxy morphology -- specifically the transition from 
an irregular, dispersion-supported system to a rotation-supported disk. 
Critically, they argue that insufficient resolution and the sensitivity of the
predicted DM evolution to  star formation, feedback, and
heating/cooling subgrid models in previous Illustris works means
that simulation predictions alone cannot reliably determine the
true $n_z$. This directly motivates an empirical measurement such as the 
one 
we present here.

 Previous attempts to measure host DM evolution from observations include 
\cite{Lin2022}, who first explored correlations between modeled host DM 
and redshift for 17 FRBs, with inconclusive results due to the small 
sample size. \cite{Sang2025} repeated the analysis with 117 localized FRBs 
and found an evolution equivalent to $n_z\sim0.9$ in our formalism 
(Equation~\ref{eq:dmz}). \cite{Kumar2025} performed a likelihood 
analysis on 65 FRBs and found $n_z=0.24\pm 1.92$, ruling out 
$n_z\gtrsim2$ at $1\,\sigma$ depending on their assumed model; however, these 
authors did not account for instrumental effects, which we show 
must be included to avoid biased results. Finally, \cite{Bernales2025} 
analyzed 12 FRBs and found $n_z=0.3\pm1.7$, with large uncertainties 
driven by the small sample size. Here we  
develop an approach that accounts for observational 
biases and apply it to a large localized FRB host sample.

\subsection{Importance of evolution constraints for FRB science}

Constraining the redshift evolution of the host DM is relevant not only 
for galaxy evolution studies, but also for the broader FRB science ecosystem  
for two main reasons.

First, the total observed FRB DM contains contributions from the Milky Way, 
cosmic structure (i.e., intergalactic medium and intervening halos), and the host galaxy. 
Isolating the cosmic  component -- which increases approximately 
linearly 
with redshift -- is the foundation of FRB-based cosmological and astrophysical 
inference. Accurate knowledge of the host DM evolution is essential to 
correctly subtract the host contribution at $z>0$ and thereby obtain an 
unbiased cosmic DM: We will show in
Section~\ref{sec:results} that our $n_z$ result shifts the inferred mean 
host DM by $\sim30\%$, an error
that propagates directly into constraints on the Hubble constant $H_0$, 
helium reionization, and the distribution of cosmic baryon density 
\citep{Mcquinn2014,Caleb2019,Macquart2020,James2022,Wang2025}, as well 
as astrophysical parameters such as galaxy feedback strength 
\citep{Baptista2023,Leung2025}, all of which are tightly linked to the 
cosmic DM \citep[][and \citealt{Glowacki2024} for a review]{Deng2014,Zhou2014,Walters2018}.

Second, a well-characterized host DM evolution  will enable 
 comparisons between FRB host populations at different cosmic epochs. 
Combined with host galaxy type (e.g., star forming or quenched) and 
properties such as star formation rate 
and stellar mass,  measurements of the evolution can shed light on 
the nature of FRB progenitors and the environments that preferentially 
produce them.

Below, we present a robust framework for measuring host DM evolution from 
observational data and provide initial constraints. Statistically strong  
results will become possible in future work with larger FRB samples, 
together with studies correlating the DM 
evolution with host galaxy type and physical properties.

Section~\ref{sec:sims} describes the 
data and surveys used in this work, and Section~\ref{sec:insight} provides 
a first look at the effect of a host DM evolution on FRB observables. 
Degeneracies and biases that may affect evolution measurements are assessed 
in Section~\ref{sec:degen}, and the results of our inference are presented 
in Section~\ref{sec:results}. We discuss the implications of our results 
and compare to previous work in Section~\ref{sec:discussion}, before 
concluding in Section~\ref{sec:conclusions}.

Unless otherwise stated, DM values are given in units of ${\rm pc\,cm^{-3}}$, 
and we quote host DM values in the rest frame of the host galaxy.

\section{Surveys and Data}\label{sec:sims}

Our calculations make use of the 
 \texttt{zDM} package\footnote{\url{https://github.com/FRBs/zdm}} 
\citep{James2021,prochaska2023,Baptista2023} within the \texttt{FRB} library \citep{Prochaska2025}.  
The main difference compared to previous \texttt{zDM} versions is the   
implementation of the  redshift  dependence for host DM described by  
Equation~\ref{eq:dmz}. We use the best fit  values  $\log\mu_{\rm host}=2.18$ and  $\log \sigma_{\rm host}=0.42$  for ${\rm DM_{host,0}}$\footnote{We note that these are not the usual $\mu$ and 
$\sigma$ parameters that characterize the log-normal distribution. For 
consistency with \cite{Hoffmann2025} and Table 1 of  
\cite{James2022}, our $\log\mu_{\rm host}$ and $\log \sigma_{\rm host}$ variables 
correspond to the $\log_{10}$ values of the mean and standard 
deviation of ${\rm DM_{host,0}}$.}, as well 
as those of other relevant parameters as reported in Figure 1 of \citealt{Hoffmann2025}  unless stated otherwise. 

We focus the analysis on the three sub-arcsecond FRB localization precision surveys/instruments with currently published FRB hosts at $z>1$: 
\begin{enumerate}
    \item The MeerTRAP transient buffer system \citep{Rajwade2024}, in coherent mode, 
    at the MeerKAT\footnote{Meer Karoo Array Telescope} radio telescope \citep{Jonas2016}. The relevant MeerTRAP FRB data is presented in \cite{pastor2026} and \cite{Caleb2026}, with one FRB host at $z=2.148$ \citep{Caleb2026}.

    \item The Deep Synoptic Array (DSA; \citealt{Kocz2019,Ravi2023}) interferometer, with the data reported in \cite{law2024,Sherman2024,Connor2024}, with one FRB host at $z=1.354$ \citep{Connor2024}.

    \item The Commensal Real-time ASKAP Fast Transients
    (CRAFT; \citealt{Bannister2019,Cho2020,Scott2023}) incoherent sum (ICS) mode survey at 900 MHz, 1.3 GHz and 1.6 GHz, with the Australian Square Kilometre Array Pathfinder (ASKAP; \citealt{Hotan2021}) telescope. This data arises from 
    \cite{Bannister2019}, \cite{Ryder2023} and \cite{Shannon2025}, as reported in \cite{Hoffmann2025}, and contains one FRB host at $z=1.016$ 
    \citep{Ryder2023}.
\end{enumerate}

\begin{figure}
\includegraphics[width=0.5\textwidth]{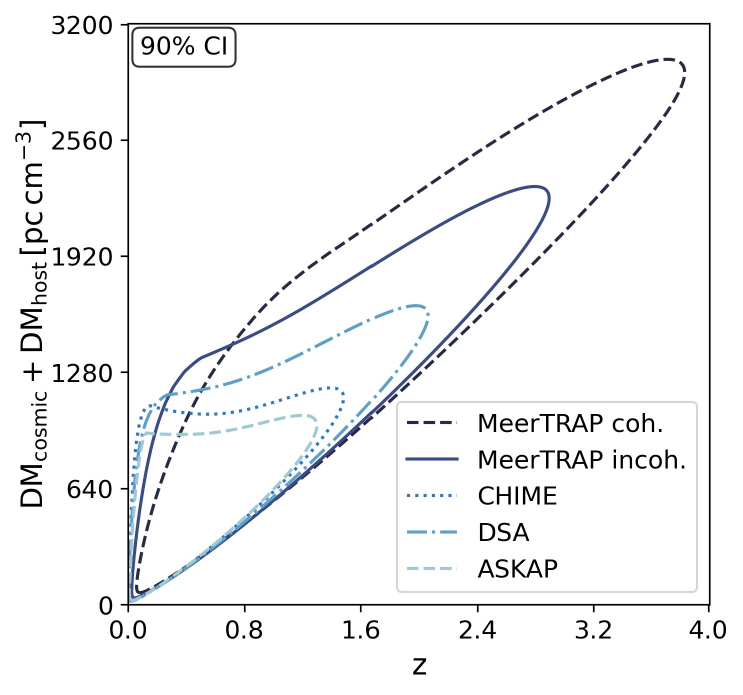}
\caption{90\% C.I. for the probability $p({\rm DM_{EG}},z)$ of detecting 
FRBs at a given extragalactic DM  and host redshift for  current sub-arcsecond FRB localization instruments.}
\label{fig:contours}
\end{figure}

\begin{figure*}\centering
\includegraphics[width=0.75\textwidth]{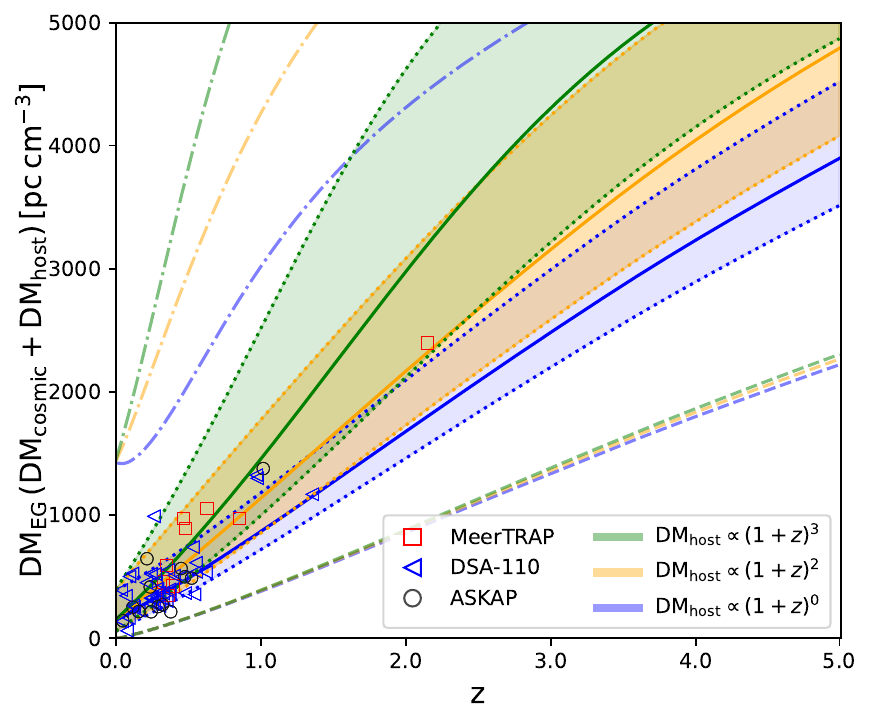}
\caption{The distribution of FRB host redshift and extragalactic DM 
for three values of the index $n_z$ that characterizes the redshift 
evolution of host DM. 
The colored solid lines and bands denote the median and the
region enclosed by the 16\% and 84\% percentiles, respectively. The dashed 
lines correspond to the $\approx0$\% percentile (DM cliff), and the 
dash-dotted to the 99\% percentile. 
The markers show MeerTRAP, DSA-110 and ASKAP FRBs with current 
identified hosts, with three at $z>1$, and one of them at 
$z>2$. The steeper the host DM evolution, the lower the host redshift at a 
given extragalactic DM. Differences between $n_z$ values are more visible 
at high redshift, making high-redshift FRBs the best indicators 
of an evolution.}
\label{fig:pzdm}
\end{figure*}

Beyond the data themselves, these surveys span a wide range of instrumental sensitivities, allowing us to assess the dependence of the results on detector sensitivity  and to (qualitatively) extend the findings to other  instruments of similar characteristics. Figure~\ref{fig:contours} shows the 90\% C.I. for the probability $p({\rm DM},z)$ of detecting 
FRBs at a given extragalactic DM (${\rm DM_{EG}} = {\rm DM_{cosmic}} + {\rm DM_{host}}$) and host redshift $z$ for all current sub-arcsecond FRB localization instruments. The MeerTRAP instrument in coherent mode 
reaches the largest FRB DM and redshift values, while ASKAP covers the 
opposite case and DSA occupies the region in between the former two. CHIME appears 
slightly above the ASKAP contours, while the incoherent mode of MeerTRAP is 
in between its coherent counterpart and DSA. The upcoming CHORD and 
DSA-2000 experiments are expected to have sensitivities comparable to 
MeerTRAP in incoherent mode \citep{connor2022}. The Bustling Universe Radio 
Survey Telescope in Taiwan (BURSTT; \citealt{Burstt2022}) also provides 
sub-arcsecond localizations but it focuses on nearby sources ($z\lesssim0.1$).

\section{Insights into DM$_{\rm \MakeLowercase{host}}\MakeLowercase{(z)}$}\label{sec:insight}

 Figure~\ref{fig:pzdm} illustrates the effect of an evolving 
host DM with redshift on the FRB DM$_{\rm EG} - z$ parameter space  for three values of the index $n_z$. We have only considered positive values 
of $n_z$ for simplicity, but negative values are also possible. 
The solid lines 
denote the median of the distributions and the bands cover the region between the 16\% and 84\% 
percentiles, where the blue color represents $n_z=0$ (no evolution). The 
data points are MeerTRAP, DSA-110 and ASKAP CRAFT/ICS FRBs with 
confirmed redshifts, where three are at $z>1$ and one at $z>2$. 
Assuming that the minimum possible host DM is the same regardless of redshift 
evolution, the minimum ${\rm DM_{EG}}$ values at a given redshift (the 0\%  
percentile; also known as DM cliff) are the same for all cases and do 
not depend on redshift evolution (dashed lines). The upper limits, however, 
are different for distinct $n_z$ values (dash-dotted lines representing 
the 99\% percentile). 

Two main 
points arise from Figure~\ref{fig:pzdm}: First, differences between evolutions 
(i.e.,  distinct $n_z$ values) become larger with increasing redshift. 
This implies that high-redshift FRBs are the strongest discriminants 
for measuring a potential galaxy 
evolution. Second, a steeper redshift evolution yields a lower host redshift  
at a fixed ${\rm DM_{EG}}$, thus narrowing the redshift distribution of the 
FRB population. We discuss these points further below.


\begin{figure*}\centering
\includegraphics[width=1.05\textwidth]{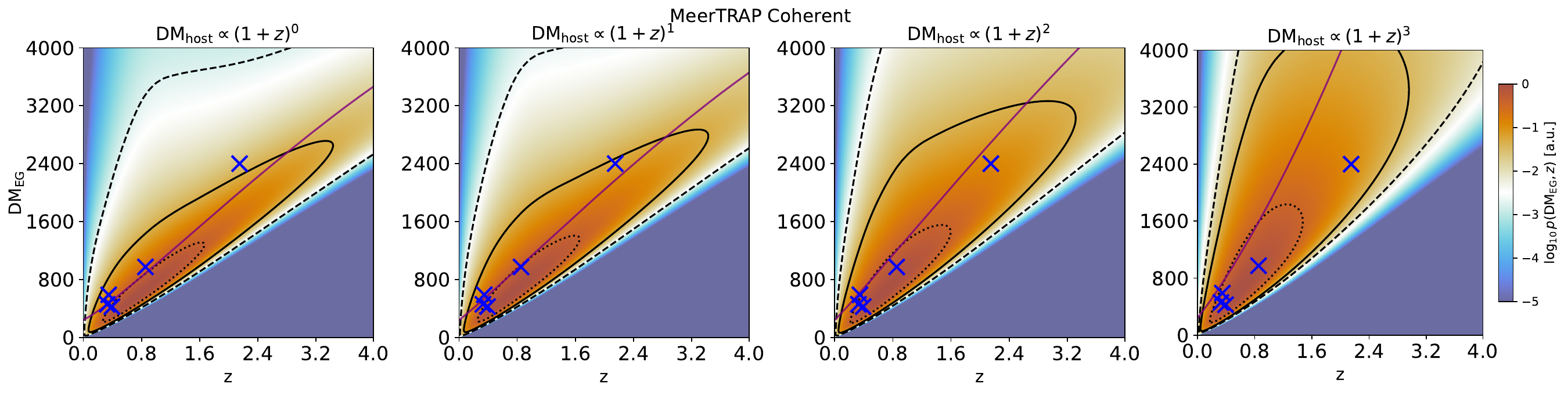}
\includegraphics[width=1.05\textwidth]{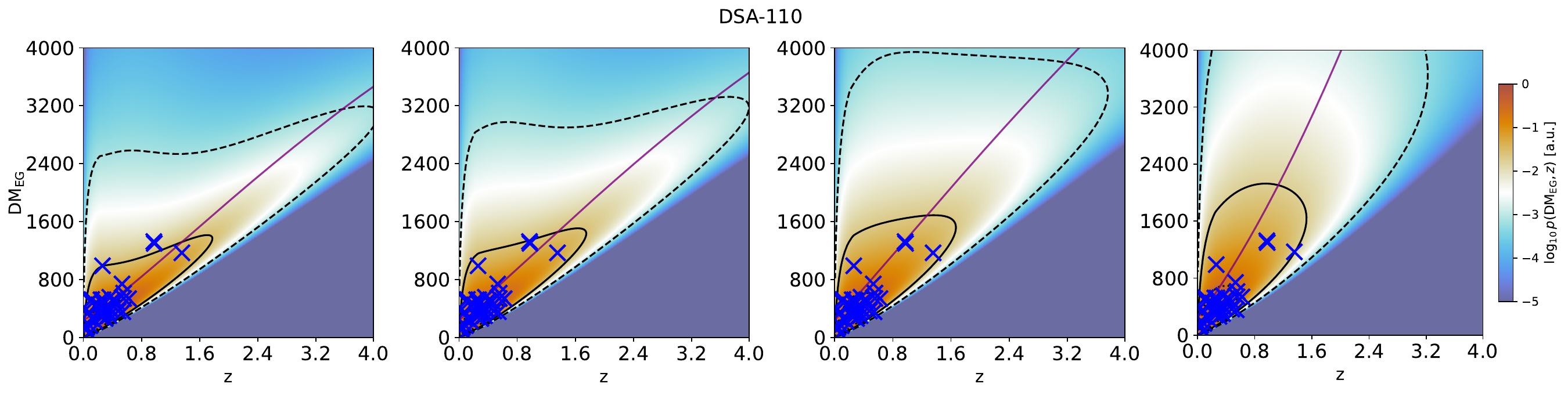}
\includegraphics[width=1.05\textwidth]{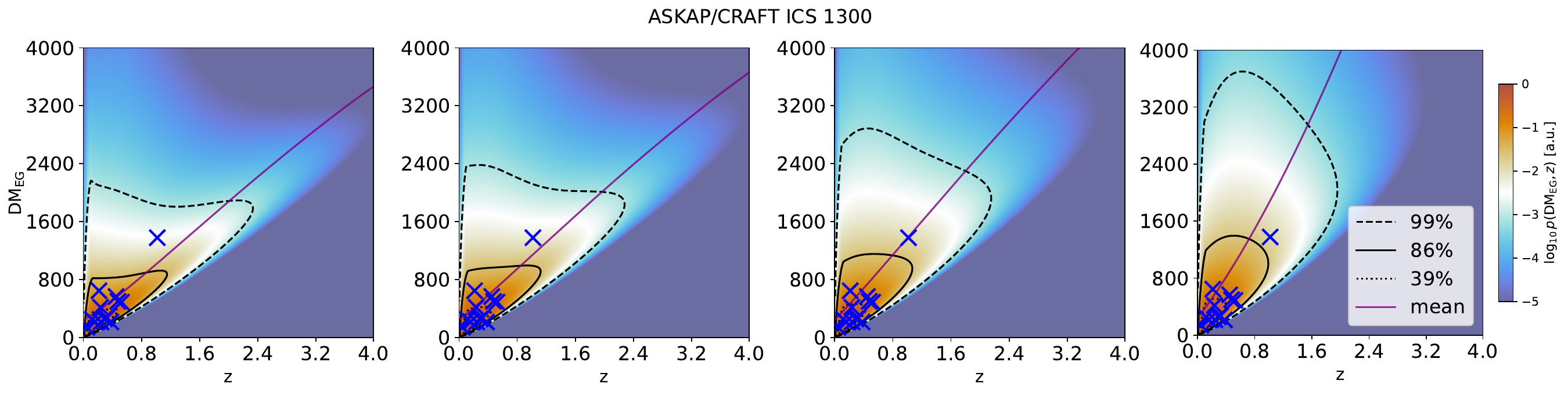}
\caption{Effect of 
 $n_z$ on the $p({\rm DM_{EG}},z)$ space of specific instruments. The leftmost column corresponds to the non-evolution case. 
The contours represent the two-dimensional 1$\sigma$, 2$\sigma$ and 
3$\sigma$ C.I., the purple lines denote the 
mean values and FRB data are plotted in blue.}
\label{fig:taudm}
\end{figure*}

\begin{figure*}\centering
\includegraphics[width=1.0\textwidth]{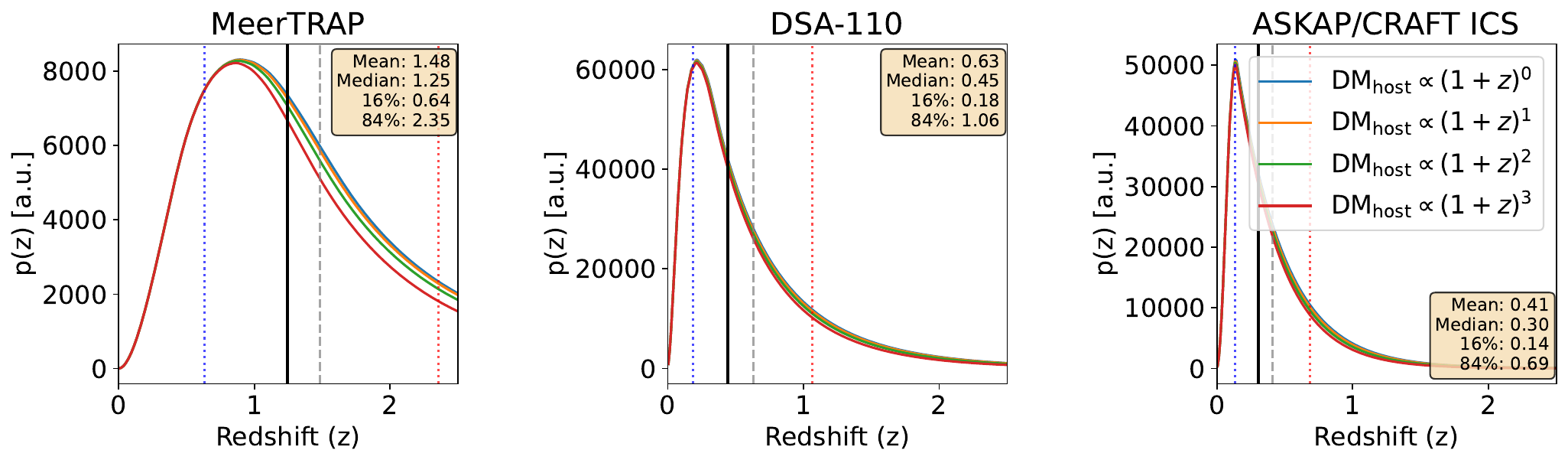}
\caption{Redshift distributions of FRBs for MeerTRAP in coherent mode, DSA and ASKAP CRAFT/ICS at 1.3 GHz, and for different host DM redshift evolutions. Steeper evolutions narrow the distributions more, most visible  for   
the MeerTRAP case (left panel). This is due to MeerTRAP highest sensitivity to high-redshift FRBs for which the differences between $n_z$ are 
larger. Mean, median, 16\% and 84\% percentiles are  shown  for the case of $n_z=1$.}
\label{fig:pz}
\end{figure*}

To quantitatively assess the existence and impact of a host evolution,  
however, the specific instrumental effects  of each experiment must 
be taken into account. This is because the \textit{observed} $p({\rm 
DM_{EG}},z)$ is different than the \textit{theoretical} one in Figure~\ref{fig:pzdm}. 
Figure~\ref{fig:taudm} illustrates the effect of 
different $n_z$ values on the observed $p({\rm DM_{EG}},z)$ space of our three fiducial 
instruments, where the leftmost column corresponds to the non-evolution case. 
The contours represent the two-dimensional 1$\sigma$, 2$\sigma$ and 
3$\sigma$ C.I. for the distributions, the purple lines denote the 
mean values and FRB data are plotted in blue. Because differences 
between evolutions are larger for high-redshift FRBs, the highest 
sensitivity (and highest fraction of high-z FRBs)  
of MeerTRAP  makes it the most 
promising instrument to test an evolution scenario, although its   
FRB detection rate is lower than that of DSA or ASKAP. The 
sensitivity difference is also 
visible in the redshift distributions shown in Figure~\ref{fig:pz}, where 
the narrowing for different $n_z$ values becomes more apparent for 
the MeerTRAP case.

\section{Parameter Degeneracies}\label{sec:degen}

The primary natural parameter degeneracy with   
the evolution index $n_z$ arises from the term $\mu_{\rm host}$  that 
characterizes the mean DM host value (Equation~\ref{eq:dmz}). 
We assess now the correlation between these two parameters and its 
impact on measuring $n_z$.

For each calculation below we use 1000 mock FRBs  sampled from the $p({\rm DM_{EG}},z)$ distribution 
of the corresponding instrument. We fix $n_z=1$ and $\log \mu_{\rm host}=2.18$, as well as $H_0=67~{\rm km\,s^{-1}\,Mpc^{-1}}$, since this 
value is consistent with cosmological results \citep{Guo2025} and with our 
best-fit result later, and fix all other parameters at the best fit 
values of 
\cite{Hoffmann2025}. Because of the grid design, we impose an upper limit on the redshift value of $z=5$, although this exact value has no impact on the results. 
Likelihoods for the FRBs in each case are then computed following 
\cite{James2021} on  
a two-dimensional grid of points covering the ranges   
$\log \mu_{\rm host} \in [1.4,\,2.4]$ and $ n_z \in [-2.2, \,5.2]$, 
with step sizes  
$\Delta \log \mu_{\rm host}= 0.1$ and $\Delta  n_z=0.37$. These ranges 
and step-size values are chosen after initial calculations and small 
variations do not alter the outcome. 

\begin{figure*}\centering
\includegraphics[width=0.33\textwidth]{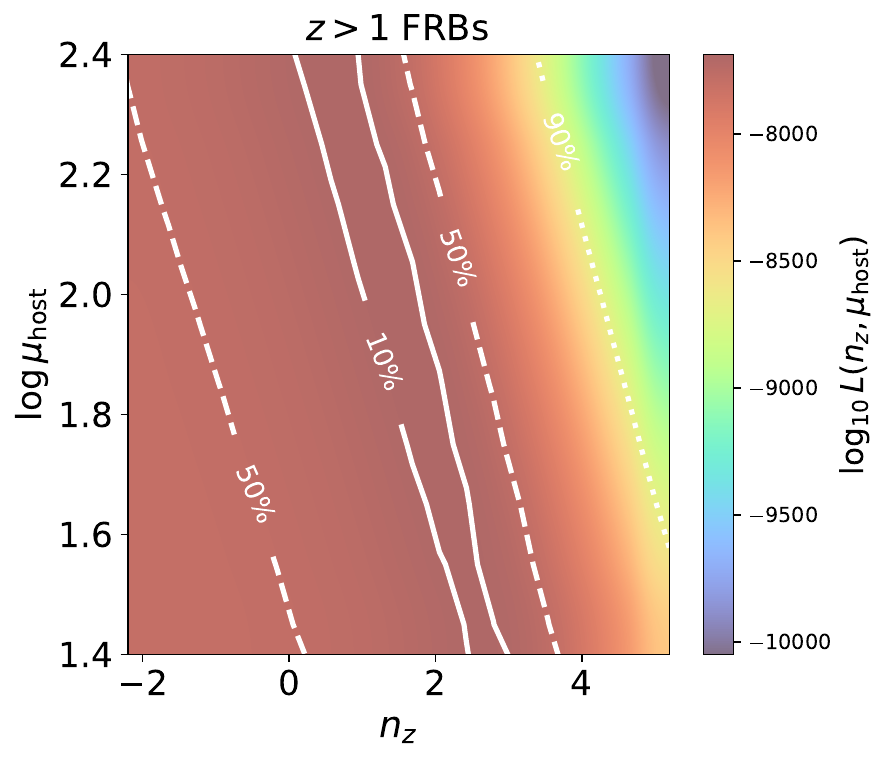}\includegraphics[width=0.33\textwidth]{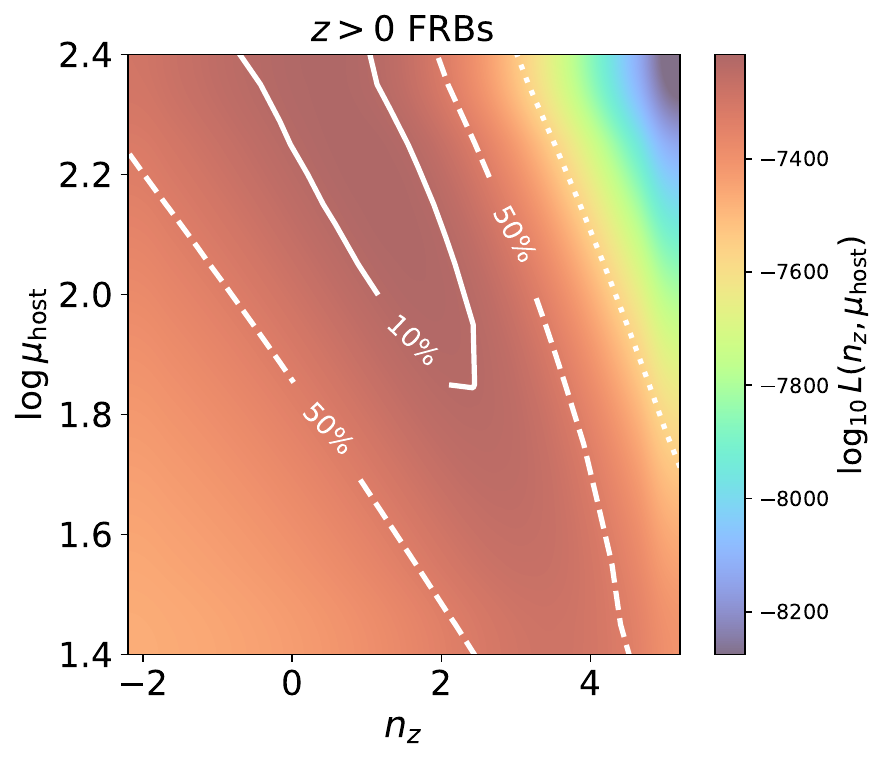}\includegraphics[width=0.33\textwidth]{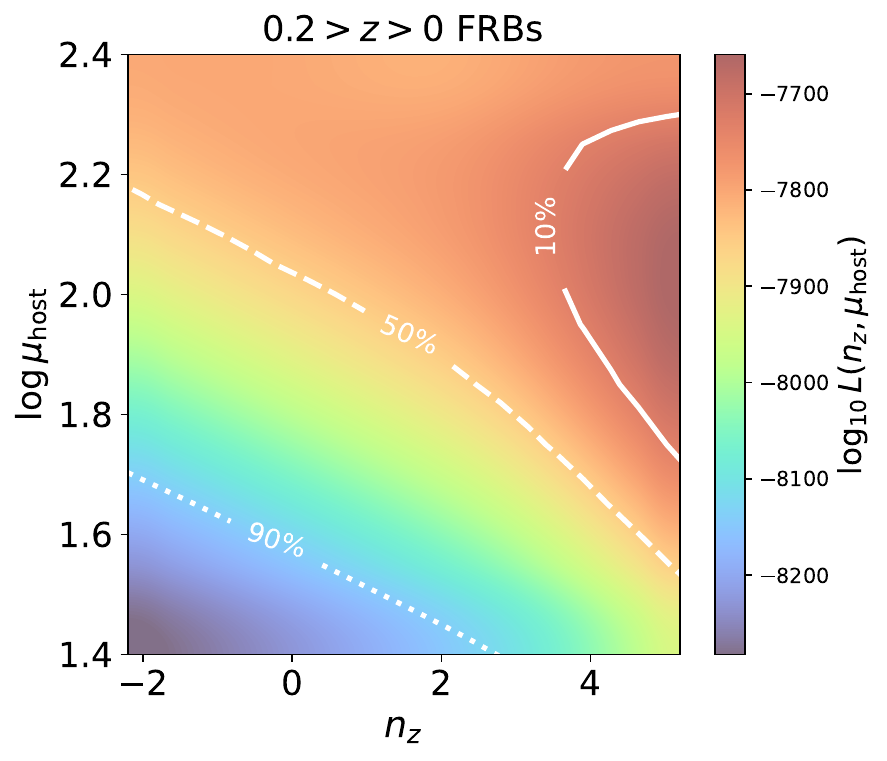}
\includegraphics[width=0.33\textwidth]{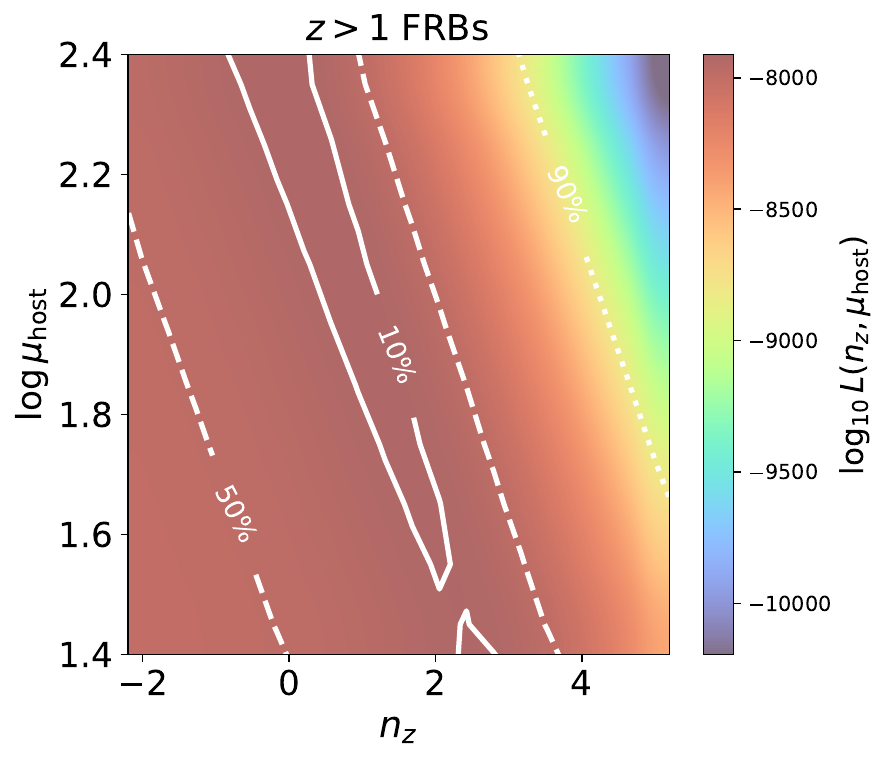}\includegraphics[width=0.33\textwidth]{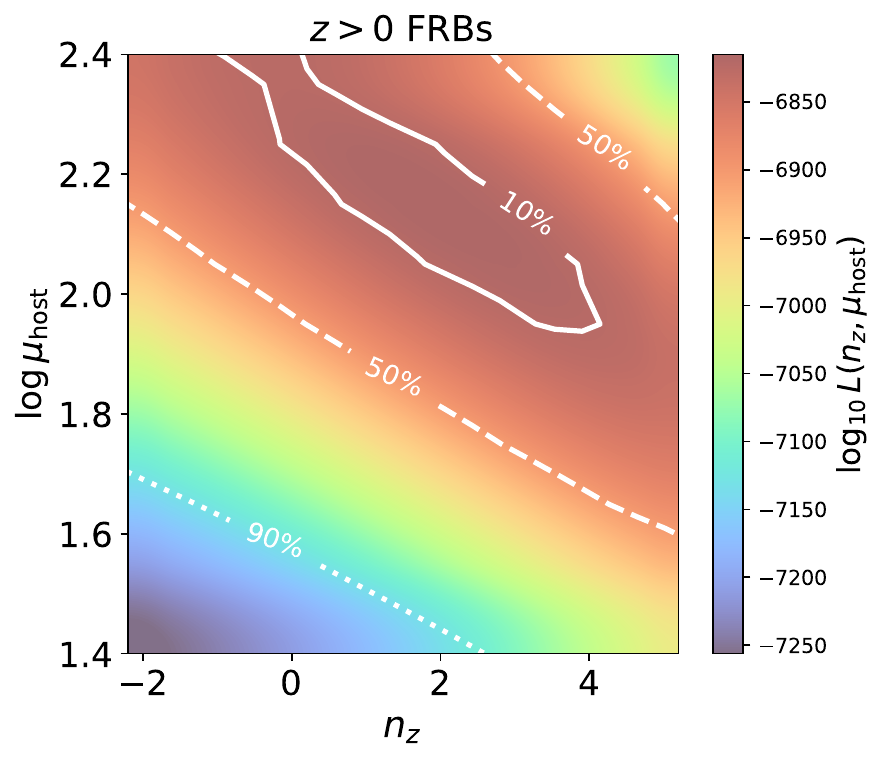}\includegraphics[width=0.33\textwidth]{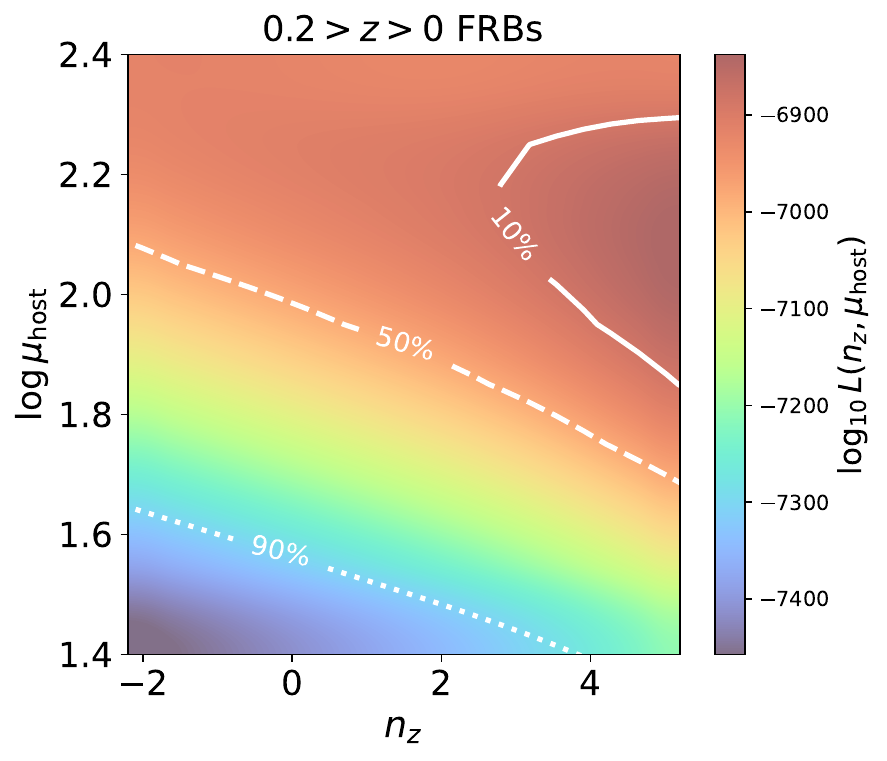}
\includegraphics[width=0.33\textwidth]{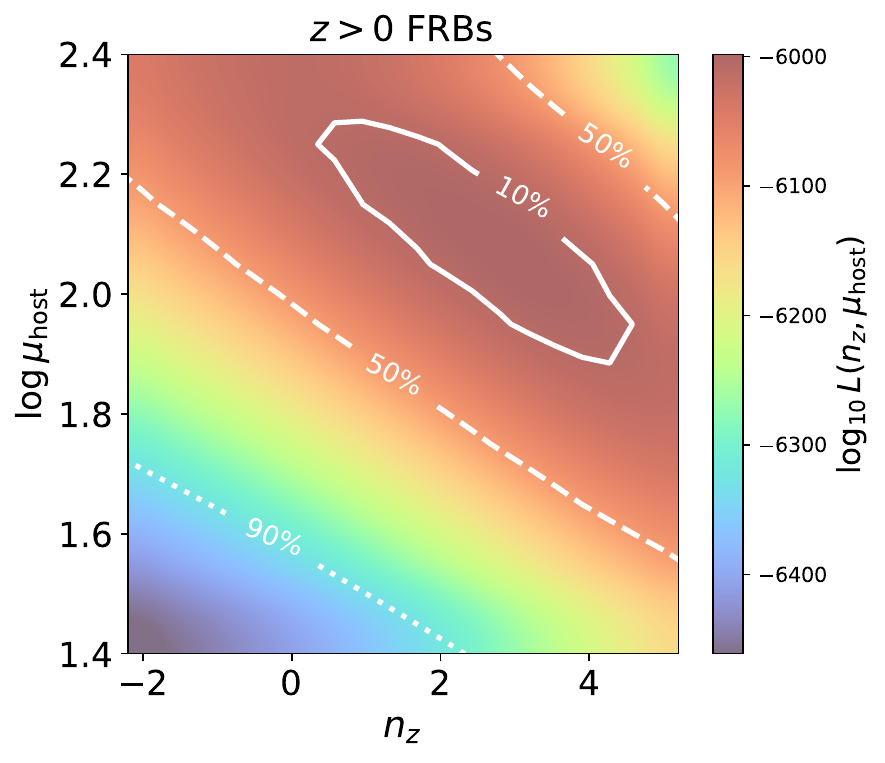}\includegraphics[width=0.33\textwidth]{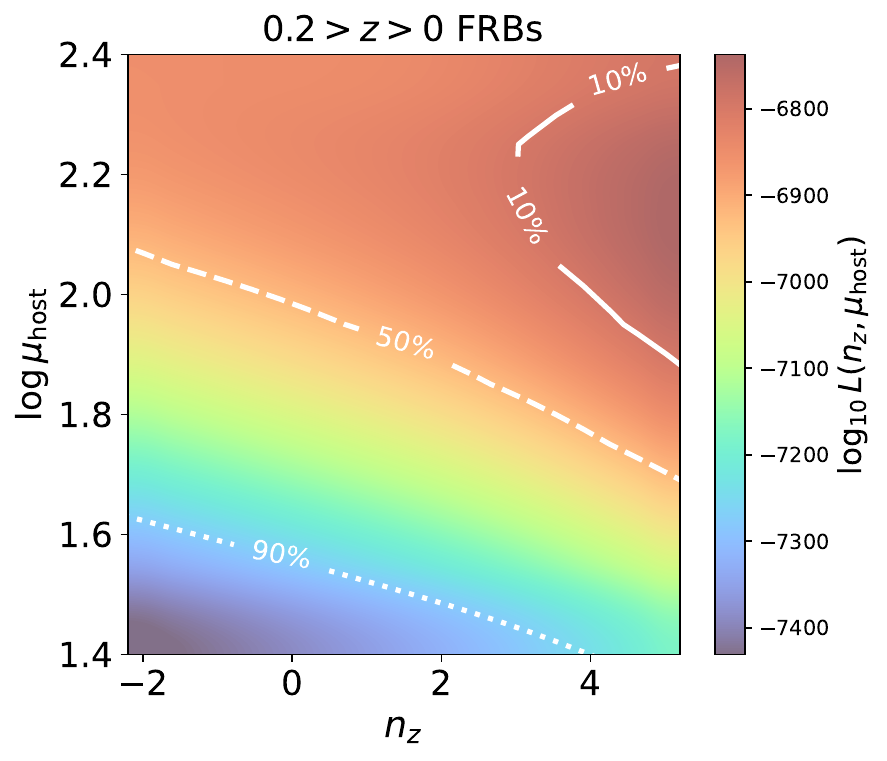}
\caption{Likelihood results for MeerTRAP (\textbf{top row}), 
DSA (\textbf{middle row}) and CRAFT (\textbf{bottom row}) FRBS, for three FRB redshift ranges (two for CRAFT given the expected small number of $z>1$ FRBs). 
High-redshift FRBs ($z>1$; leftmost column in the top two rows) show  
strong (weak) constraining power on $n_z$ ($\mu_{\rm host}$), while the 
opposite is true for 
low-redshift FRBs ($0.2>z>0$; rightmost panels). The degeneracy between 
$\mu_{\rm host}$ and $n_z$ is most notable when the sample includes high-redshift FRBs (middle and left columns).}
\label{fig:degen}
\end{figure*}

Figure~\ref{fig:degen} shows the likelihood results for MeerTRAP (top row), 
DSA (middle row) and CRAFT (bottom row) FRBS, for three FRB redshift windows (two for CRAFT given the expected small number of FRBs at $z>1$). 
High-redshift FRBs ($z>1$; leftmost column in the top two rows) show  
strong (weak) constraining power on $n_z$ ($\mu_{\rm host}$), while the 
opposite is true for 
low-redshift FRBs ($0.2>z>0$; rightmost panels). Furthermore, the 
degeneracy between $\mu_{\rm host}$ and $n_z$ is most notable when the 
sample includes high-redshift FRBs (middle and left columns): This is 
because, for high-redshift FRBs, $n_z$ can easily replicate a change in 
DM (at a fix $z$) induced by a variation in $\mu_{\rm host}$ by changing the 
curvature of the redshift evolution. However, the required change in 
$n_z$ (curvature) to modify DM tends to infinity when approaching $z=0$, 
and so the degeneracy diminishes for low-redshift FRBs.  
The middle column in Figure~\ref{fig:degen} further shows that the degree of 
degeneracy depends on the instrument. Because MeerTRAP has a larger fraction 
of high-redshift FRBs compared to DSA and CRAFT, the anti-correlation 
between $\mu_{\rm host}$ and $n_z$ is steeper than that of the other two 
instruments. 

In summary, MeerTRAP appears as the most sensitive instrument to measure 
$n_z$ given its largest fraction of high-redshift FRBs compared to other 
experiments. However, using both high- and low-redshift FRBs gives the best joint 
constraints on $\mu_{\rm host}$ and $n_z$, which may be best achieved 
by combining data from different instruments. Trying to infer $n_z$ with 
low-redshift FRBs alone  may lead to biased results. 

Finally, Figure~\ref{fig:sig} in Appendix~\ref{sec:applike} shows 
similar likelihood calculations but replacing $\mu_{\rm host}$ by 
$\sigma_{\rm host}$. This is motivated by the 
$\sigma_{\rm host}$ relation with host DM and because this parameter, 
like  $\mu_{\rm host}$, was found to be degenerate with $H_0$ in   
\cite{James2022}. No significant correlations appear between 
$\sigma_{\rm host}$ and $n_z$ in our test, so we exclude this parameter in 
subsequent calculations.

\section{Measuring DM$_{\rm \MakeLowercase{host}}\MakeLowercase{(z)}$} \label{sec:results}

We perform an inference analysis on observational data to quantify $n_z$ below, 
and then explore the number of FRBs required to reduce the uncertainties.

\subsection{$n_z$ from DSA and CRAFT}

We now infer $n_z$ via a Markov Chain Monte Carlo (MCMC) analysis of the 
DSA and ASKAP/CRAFT ICS FRBs. We exclude MeerTRAP coherent-mode FRBs 
given their current small sample size. The DSA and CRAFT samples contain 
47 (39) and 43 (30) FRBs (with measured host redshift), respectively, 
where the CRAFT ICS FRBs arise from the three, 900 MHz, 1300 MHz and 
1600 MHz, frequency bands.

We consider $\mu_{\rm host}$, the 
cosmological Hubble constant $H_0$ and the astrophysical feedback/fluctuation parameter $F$  as free parameters to explore further potential degeneracies.  
This is motivated by the work of \cite{James2022} and \cite{Baptista2023} 
who found that $H_0$ is strongly degenerate with 
$\mu_{\rm host}$ and $F$, respectively, and we just  demonstrated  
that $\mu_{\rm host}$ is degenerate with $n_z$. As noted above for 
$\sigma_{\rm host}$,  degeneracies 
with additional parameters are expected to be less important and we do 
not account for them here, but future work may take them into account 
as we discuss in Section~\ref{sec:discussion}.

Owing to the computational cost of MCMC calculations, we here use a discrete 
likelihood approach, where the likelihoods arise from a finite number of 
discrete prior values. In practice, this entails the continuous sampling 
of the prior parameter space, but the sampled priors are then rounded 
to the closest discrete values and the corresponding (precomputed) 
likelihoods are adopted. We have tested that the low dimensionality of our 
problem and the  step sizes for the discrete priors yield valid results. 
Table~\ref{tab:priors} quotes the values for the discrete priors, chosen to 
cover the range of interest for our results after initial calculations. In 
particular, the lower $H_0$ bound corresponds to the current CMB 1$\sigma$ 
lower limit \citep{Planck2020}, and we discuss the prior ranges further in 
Section~\ref{sec:discussion}. 
For the Markov chain we use \texttt{emcee} \citep{emcee2013}, 
with 64 walkers and 9000 steps after a burn-in stage of 1000 steps, and 
initial guesses covering the prior ranges uniformly.  

\begin{table}[]
    \begin{tabular}{cccc}
         Parameter &Min &Max &Step \\
         \hline 
         $\log \mu_{\rm host}$ &1.5 &2.4 &0.1\\
         $n_z$ &-2.2 &5.2 &0.2\\
         $H_0$ &67 &71 &0.5\\
         $\log F$ &-1.0 &-0.2 &0.2\\
    \end{tabular}
    \caption{Discrete prior values for the free parameters  in the MCMC calculation.}
    \label{tab:priors}
\end{table}

\begin{figure*}\centering 
\includegraphics[width=0.54\textwidth]{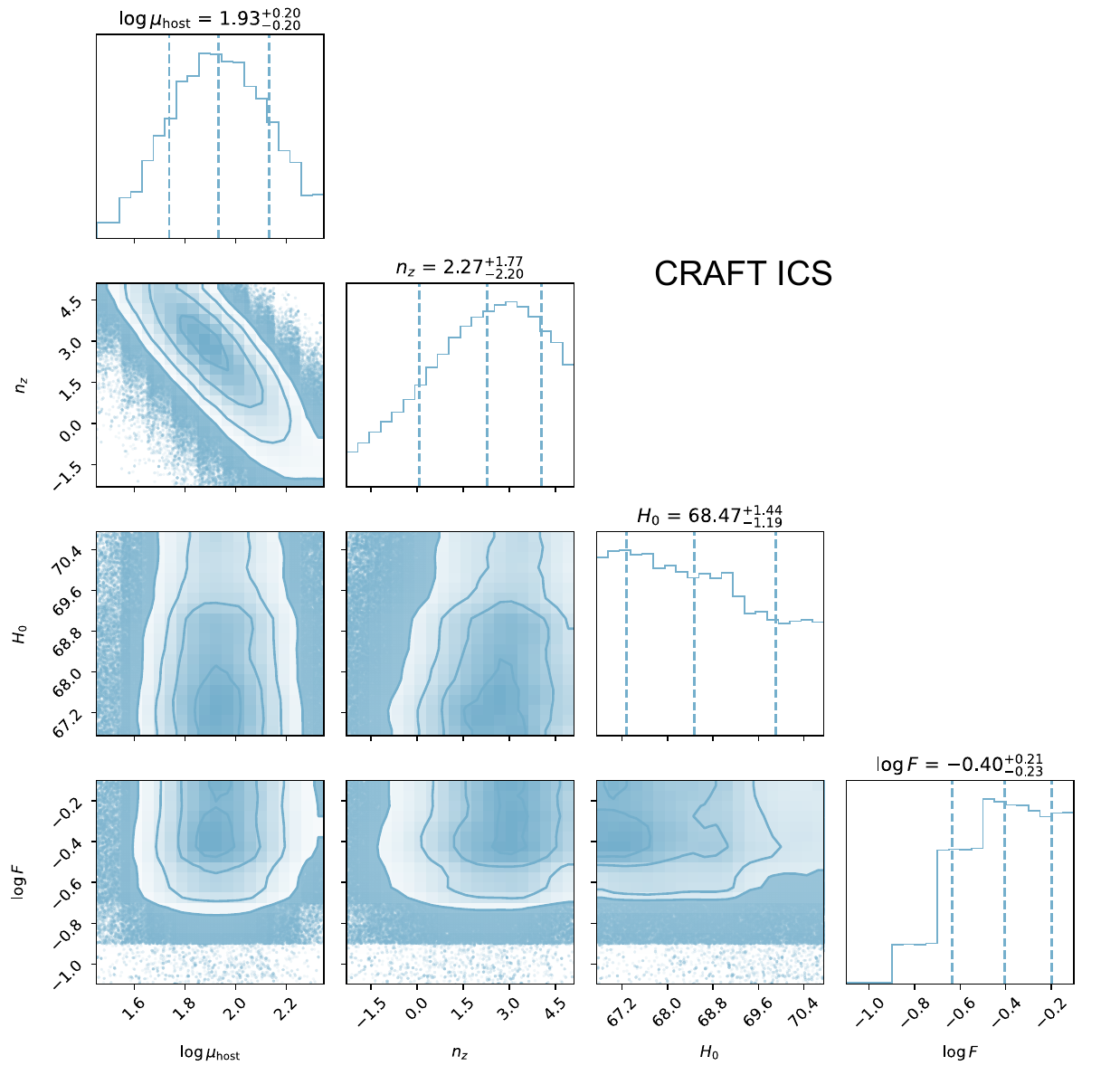}\includegraphics[width=0.54\textwidth]{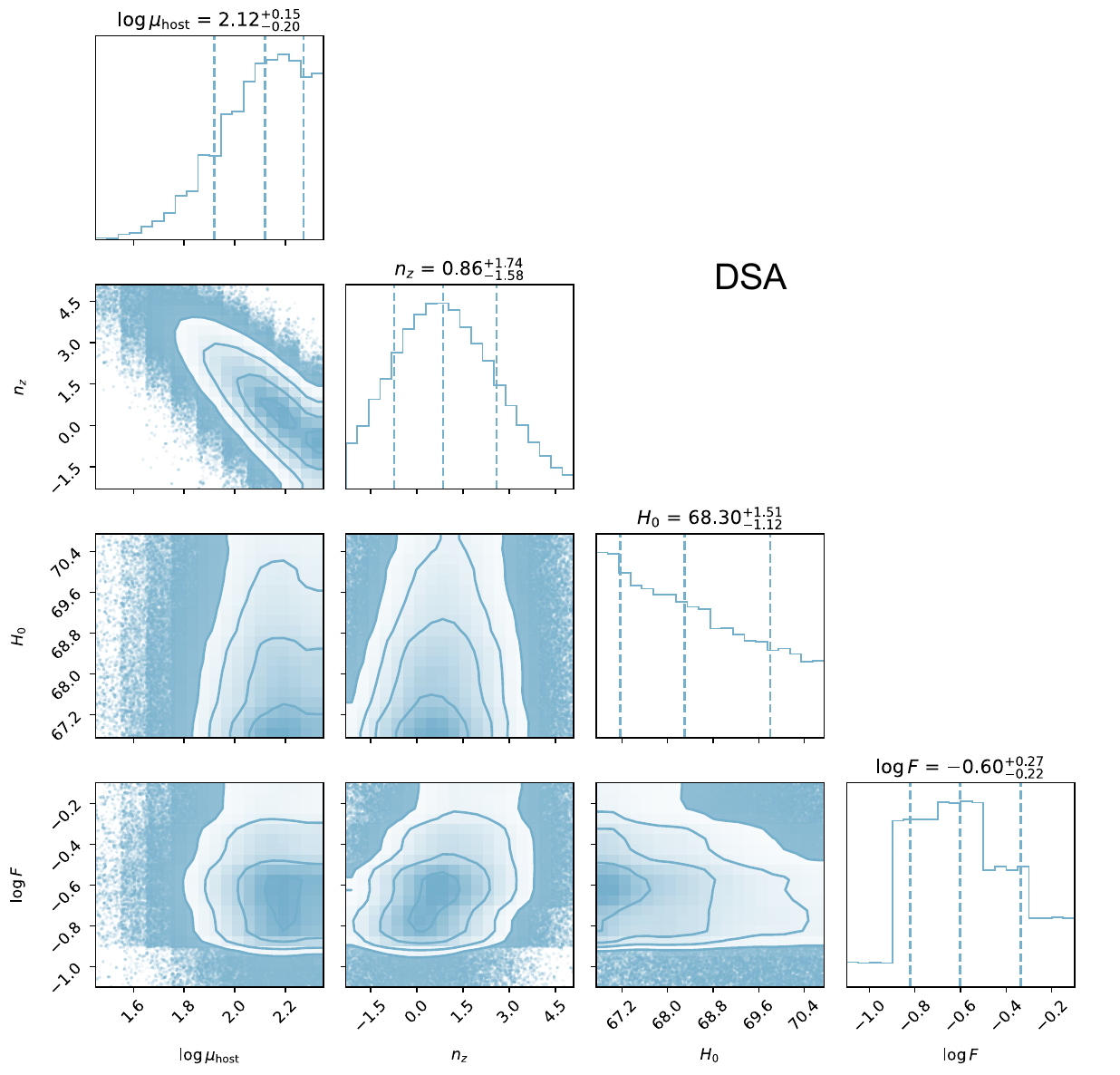}
\includegraphics[width=0.825\textwidth]{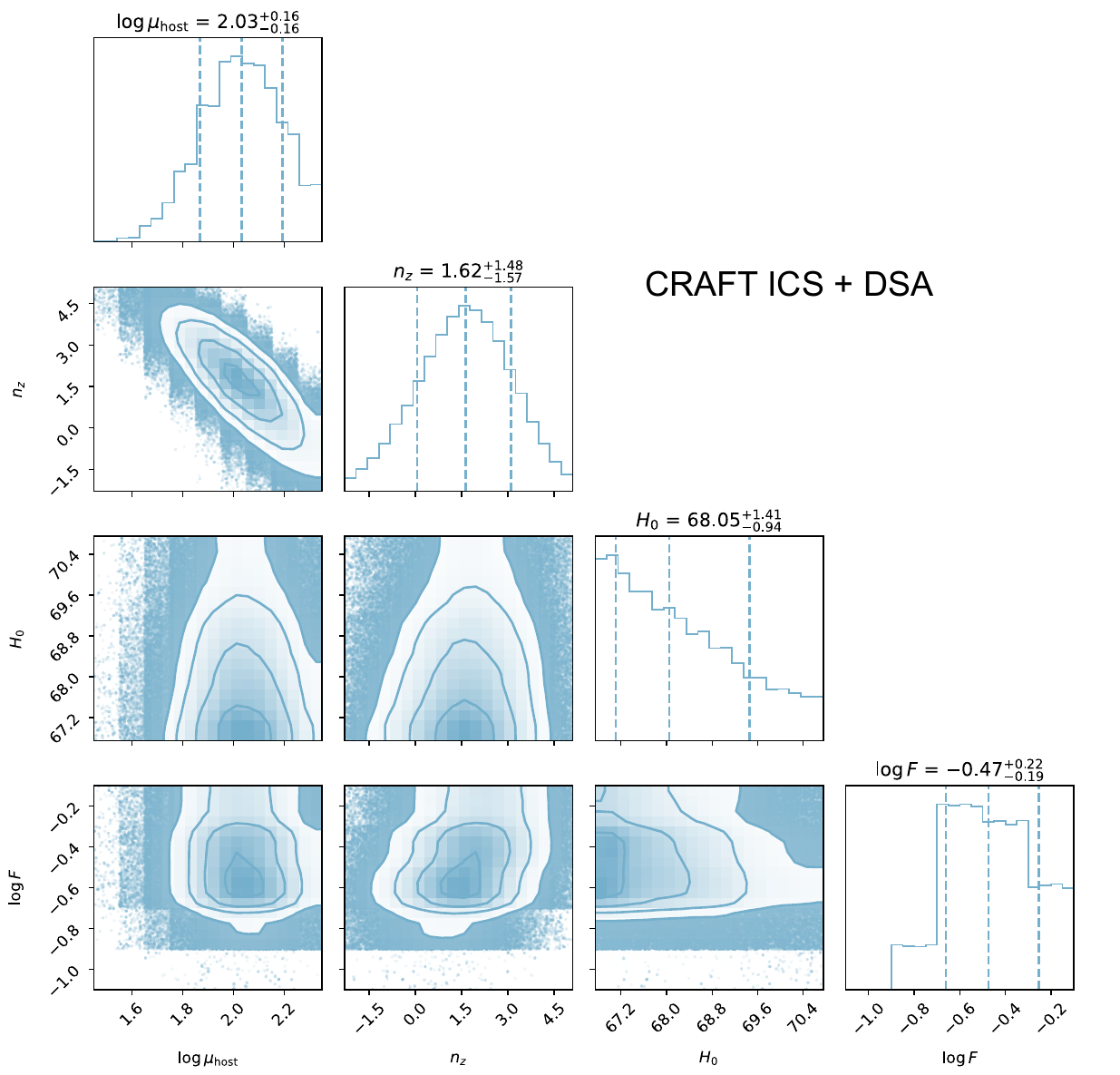}
\caption{Inference results for 43 (30 with redshift) ASKAP/CRAFT ICS FRBs (top left), 47 (39) DSA (top 
right) and the two joint datasets (90 FRBs, 69 with redshift; bottom). Correlations between $n_z$ and 
other parameters are visible, most notably with $\mu_{\rm host}$. A host DM 
evolution ($n_z>0$) is favored in all cases.}
\label{fig:corner}
\end{figure*}

Figure~\ref{fig:corner} shows the MCMC results. The top left and right 
plots correspond to the ASKAP/CRAFT ICS and DSA FRBs, respectively, 
and the bottom plot is our main result from the two joint datasets. 
The joint dataset rules out a non-evolution ($n_z=0$) scenario at 
1\,$\sigma$, where $n_z=1.62^{+1.48}_{-1.57}$.  The CRAFT and DSA posteriors 
peak at $n_z\approx3$ and $n_z\approx0.85$, respectively, both also favoring the 
existence of an evolution, although with large uncertainties compatible with 
$n_z\sim0$. The large CRAFT $n_z$ value is driven by the single $z>1$ 
FRB in the sample, which shows a notably high ${\rm DM_{EG}}$ value at its 
redshift\footnote{We do not attempt to quantify the impact of this single FRB by 
removing it from the sample, since it is the only $z>1$ CRAFT FRB and its 
removal would leave the CRAFT $n_z$ posterior essentially unconstrained 
at high redshift, where most of the discriminating power between 
evolution scenarios lies (Section~\ref{sec:insight}). A meaningful 
robustness test will require a larger sample of high-redshift CRAFT 
FRBs, which we defer to future work.}. Contrarily, 
the $z>1$ DSA FRB is below the mean $z - {\rm DM_{EG}}$ relation 
for any $n_z>0$ case, thus favoring low $n_z$ values. Combining 
datasets (larger sample size) thus reduces the impact of particular individual 
FRBs and yields more reliable results. 

The expected strong correlations between $n_z$ and $\mu_{\rm host}$ are 
visible in all corner plots of Figure~\ref{fig:corner}, and correlations 
of $n_z$ with $H_0$ and $F$ are most apparent 
in the DSA and joint samples.  A host DM evolution 
$n_z=1.62$ yields a $\log \mu_{\rm host} = 2.03$,  a $\sim30\%$ 
lower (but within 1\,$\sigma$) than the value $\log \mu_{\rm host} = 2.18$ 
found by \cite{Hoffmann2025}, but consistent with recent 
work inferring DM$_{\rm host}$ through other approaches \citep{Ilya2024,Bernales2025}. Despite the coarse discrete values of $F$, 
our results are also in agreement with the findings 
of $\log F = -0.49$ by \cite{Baptista2023}. The posteriors of $H_0$ peak at the low CMB limit 
$H_0= 67{\rm \, km\,s^{-1}\,Mpc^{-1}}$, and we discuss the effect of  
allowing a lower limit $H_0= 65{\rm \, km\,s^{-1}\,Mpc^{-1}}$ 
in Section~\ref{sec:discussion}. 

Overall, our results favor the scenario of an evolving host DM 
with redshift and rule out the non-evolution case ($n_z=0$) at 1\,$\sigma$, 
with $n_z=1.62^{+1.48}_{-1.57}$. 
The existence of such an evolution and the $n_z$ correlations with  
$\mu_{\rm host}$, $H_0$ and $F$ stress the need to include $n_z$ as 
a parameter (or another characterization of the host DM evolution) to 
avoid biased results 
in inference analyses where host DM 
plays a role. 

\subsection{How many FRBs to constrain $n_z$?}

The $n_z$ uncertainties in our inference calculations   
can be approximated to first order as having a statistical origin and thus scale 
as $1/\sqrt{N_{\rm FRB}}$, where $N_{\rm FRB}$ is the number of FRBs. 
Considering our results and that 25 FRBs without redshift have the same 
weight as one with redshift \citep{James2022}, 100 CRAFT (DSA) FRBs with localized 
redshifts would yield $n_z$ uncertainties of $\sim 1.1$ ($\sim1$). And about 350 CRAFT (300 
DSA) FRBs are required to obtain $n_z$ uncertainties of $\sim 0.7$ ($\sim 0.6$). 

For MeerTRAP in coherent mode, we follow the approach of \cite{James2022} and 
\cite{Baptista2023} and create a mock dataset of 100 FRBs by sampling 
the $p({\rm DM_{EG}},z)$ distribution of this instrument with parameters 
$n_z=1.62$, $\log \mu_{\rm host}=2.03$ and $H_0=67$ (our best-fit values), and fixing all 
others at the best-fit values of \cite{Hoffmann2025}. We then repeat the previous 
inference analysis on this dataset but with $F$ fixed to speed up the 
calculations. Given the moderate correlation of $n_z$ with $F$, 
we expect this change to have little effect on the estimate of the uncertainty. 
This exercise  yields an uncertainty on $n_z$ of 
$\approx0.7-0.8$, similar to CRAFT and DSA but with about one third of the data. 

Finally, the rapidly-increasing number of low-redshift FRBs with identified 
host redshifts is expected to soon yield measurements of  
$\mu_{\rm host}$ at high precision. To account for this, we repeat the previous MeerTRAP 
calculation but now assigning a Gaussian prior to $\mu_{\rm host}$, 
with mean $\log \mu_{\rm host}=2.03$ and standard deviation 
$\sigma_{\log \mu_{\rm host}}=0.1$ (a conservative 
value considering that \cite{Hoffmann2025} already report  
$\sigma_{\log \mu_{\rm host}}\sim0.12$). This constraint further 
reduces the uncertainties in $n_z$ by $\approx 15\%$.

\section{Impact of a host DM evolution}\label{sec:impact}

We organize the following discussion around two themes:
implications for the FRB population itself (points $1-2$), and
implications for FRB-based inference and galaxy evolution science
(points $3-5$). Throughout, we take $n_z=1.62$ at face value for
illustration; the large uncertainties on this value mean that
quantitative conclusions should be treated as indicative rather
than definitive.

\begin{itemize}[leftmargin=*]
    \item {Probing an undetected FRB population with large host DM at low redshift};  It has been noted that several high redshift FRBs show 
    notably larger  (excess) rest-frame host DM values than their low-redshift 
    counterparts \citep[see, e.g.,][]{Ryder2023,Caleb2026}. A confirmation of this host DM 
    excess together with a lack of host evolution 
    ($n_z=0$, or an evolution insufficient to explain such high values) may 
    suggest the existence of an undetected similar FRB 
    population at low redshift. Large-DM$_{\rm host}$ FRBs at low redshift 
    may appear undetected due to the instrumental bias against high DM and 
    high scattering (width) FRBs \citep[Equations $2 -4$ in][see also 
    \citealt{Hoffmann2024,James2025}]{Masribas2026}. For high-redshift FRBs, 
    instead, a detection is possible because the observed host scattering and DM  
    values are reduced by factors $(1+z)^{-3}$ and $(1+z)^{-1}$, respectively, 
    which brings them down into the FRB detectability window. The existence of 
    a large-DM$_{\rm host}$ FRB population at low redshift would have deep 
    implications for our understanding 
    of FRBs, and it is therefore of high interest. The confirmation 
    of an evolution with $n_z\approx1.62$, however, would provide an explanation 
    for the high host DM values and would render the undetected low-redshift population unlikely.

    \item Shedding light onto the origins of FRBs; A host DM evolution 
    matching that of star-forming galaxies may be suggestive of young progenitor channels, such as magnetars and supernovae. Contrarily, an evolution resembling that of passive/quenched galaxies with old stellar 
    populations may be an indicator of delayed progenitor mechanisms 
     \citep{Bochenek2021}. Indeed, simulation work by \cite{Mo2023}, 
     \cite{Kovacs2024} and \cite{Orr2024} suggests different host DM 
    evolutions between young star-forming and old quenched galaxies. It is 
    also possible that different FRB types (e.g., repeaters and non-repeaters) 
    have different evolutions, which may provide additional information 
    about their origins. A large value of $n_z=1.62$ may be suggestive of 
    star-forming systems, thus favoring young progenitor channels, 
    although simulation results show large variations for the two galaxy 
    types and more work is required (also see Section~\ref{sec:intro} for 
    further discussion on simulation work). 

    \item Revealing the driver of the density evolution in the galaxy-halo system;  Assuming that the FRB host population 
    is dominated by star forming galaxies,  
    as currently suggested \citep[e.g.,][]{Gordon2023}, and that the 
    ionized gas environment evolves as $(1+z)^{-1}$ (expected for halos and 
    star-forming galaxies; Figure~\ref{fig:cart}), $n_z=1.62$ gives an electron 
    density evolution of $n_e \propto (1+z)^{\approx 2.6}$. This value 
    is in agreement 
    with the cosmic star formation rate evolution between $z\sim 0-2$, suggesting 
    that star formation may govern the overall density evolution. This 
    contrasts with the cases of cosmic expansion with $n_e \propto (1+z)^{3}$ 
    and \ion{H}{2} gas with $n_e \propto (1+z)^{2}$. It may also indicate that 
    the DIG phase is the dominant tracer of the ionized gas in the whole system 
    since its evolution matches the one found here. 
    We stress that the discussion here 
    is purely for illustration; The aforementioned decomposition assumes a fixed host
    size scaling and a specific host population, while 
    ${\rm DM_{host}}$ encodes the \textit{joint} evolution of gas
    density and  path length, which current
    data cannot disentangle. Robust conclusions on the dominant ionized phase and the driver of 
the density evolution will require tighter constraints on $n_z$ 
from future, larger FRB samples, combined with a detailed comparison 
to galaxy evolution observations and simulations.

    \item Inferring more accurate (unbiased) cosmic parameters; 
Beyond the galaxy evolution context, a better characterization 
of the host DM evolution also yields more accurate estimates of 
the cosmic DM contribution and, in turn, the cosmological and 
astrophysical parameters that depend on it.
    This affects empirical analyses modeling each DM component 
    separately \citep[e.g.,][]{Ilya2024, Ilya2026,Masribas2025,Kahinga2026}, 
    as well as statistical (inference) analyses where correlations between  
    host DM evolution and  cosmological or astrophysical parameters may 
    exist \citep[e.g.,][]{Wu2022,James2022}. Most notably, the result  
    $n_z=1.62$ yields a value $\log\mu_{\rm host}=2.03$, lower than the $\log\mu_{\rm host}=2.18$ by \cite{Hoffmann2025} who does not consider 
    a host DM evolution. Other parameters appear to be less affected, although 
    this will become important when highly-precise measurements are required.

\begin{figure*}\centering
\includegraphics[width=0.99\textwidth]{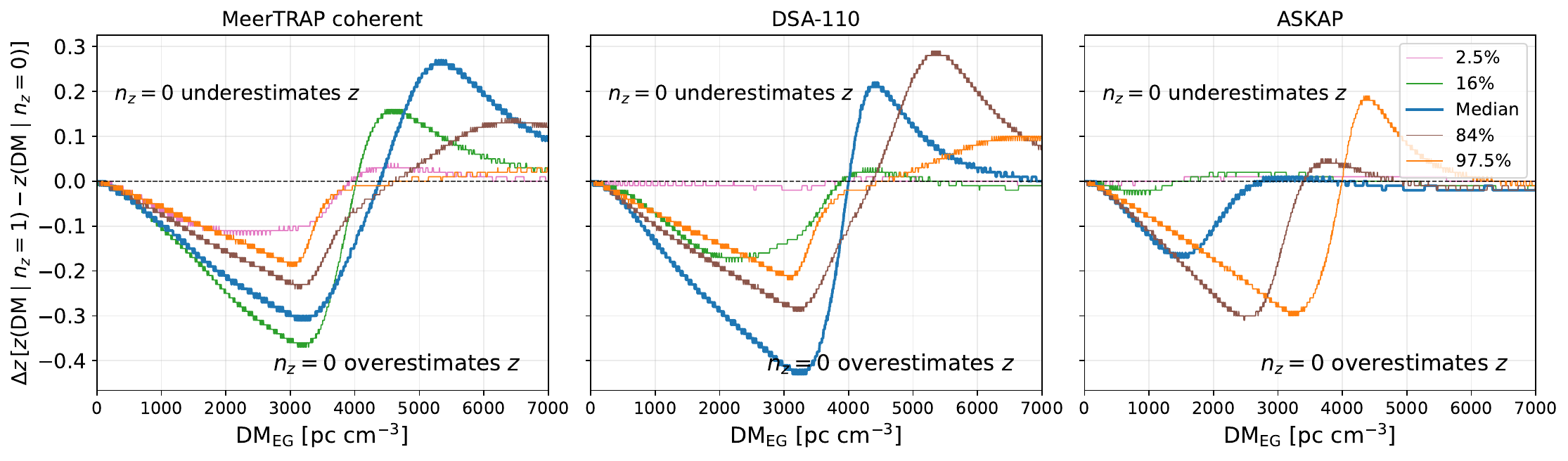}
\caption{DM$_{\rm EG}$-based FRB redshift differences of the non-evolving host 
DM case ($n_z=0$) compared to a  
$n_z=1$ host DM evolution. Colors denote the percentiles of the estimated 
conditional redshift distribution at a given DM$_{\rm EG}$, i.e, $p(z|{\rm DM_{EG}})$. Overall, ignoring the host DM evolution $n_z=1$ yields 
overestimated redshift values up to about DM$_{\rm EG}\sim3000-4000{\, \rm pc\, cm^{-3}}$, where the effect reverses due to the shape of the joint  $p({\rm DM_{EG}},z)$ distribution.}
\label{fig:nzsub}
\end{figure*}

    \item Improving FRB redshift estimates. Knowledge of the host 
    DM evolution yields more accurate FRB 
    redshifts based on DM$_{\rm EG}$, i.e., via the conditional 
    $p(z|{\rm DM_{EG}})$ distribution. Figure~\ref{fig:nzsub} shows the  
    comparison between redshifts estimated from DM$_{\rm EG}$ when 
    considering a conservative host DM evolution $n_z=1$ and no evolution ($n_z=0$), 
    for our three fiducial 
    surveys. Overall, assuming no evolution for an actual value of $n_z=1$ 
    overestimates the redshift by $\Delta z \approx 0 - 0.5$ up to 
    DM$_{\rm EG}\sim3000-4000{\, \rm pc\, cm^{-3}}$, the exact values 
    depending on the survey considered and the specific percentile of the 
    $p(z|{\rm DM_{EG}})$ distribution. This is expected since the existence 
    of an evolution generally yields lower redshifts and narrower redshift 
    distributions 
    (Figure~\ref{fig:pz}). At higher DM$_{\rm EG}$ values, however, the shape 
    of the joint $p({\rm DM_{EG}},z)$ distributions results in the reversal 
    of this effect, yielding the underestimation of redshift values when a 
    non-evolution is assumed. The observed differences would be larger for the value $n_z=1.62$.  A known host evolution will allow  
    correcting (accounting for) this effect and estimating more precise  
    redshifts.

\end{itemize}

\section{Discussion}\label{sec:discussion}

\subsection{On Constraints and Parameters}

Our work shows that FRB host DM evolves with redshift as ${\rm DM_{host}}\propto 
(1+z)^{n_z}$ with $n_z=1.62^{+1.48}_{-1.57}$~, where we have considered 
the minimum allowed value of the free parameter $H_0= 67{\rm ~ km\,s^{-1}\,Mpc^{-1}}$. 
This bound  corresponds to the 1\,$\sigma$ lower limit from CMB experiments 
\citep[\citealt{Planck2020}; and $\approx2.5\,\sigma$ considering DESI results;][]{Desi2024}, and 
other favored values are even higher \citep[$H_0\gtrsim 72{\rm \, km\,s^{-1}\,Mpc^{-1}}$ at 1$\sigma$ from the SHOES collaboration;][]{shoes2022}. However, because 
the posterior distributions of $H_0$ peak at the lower limit in our inference 
analyses (Figure~\ref{fig:corner}), we repeated the calculations  allowing a lower bound of $H_0= 65{\rm \, km\,s^{-1}\,Mpc^{-1}}$ for comparison. 
Results with this lower limit still preferred the minimum $H_0$ value, 
driving the joint dataset value to $n_z\sim 1$, lower than our main 
result but within 1$\sigma$. This preference for low(est)  
$H_0$ values was also found by \citealt{Hoffmann2025} (their Appendix 1.1), 
who traced it back to degeneracies and the modeling of the Milky Way DM 
uncertainty. The same reasoning should apply to our data and, therefore, we 
conclude that it 
is not a physical effect biasing our $n_z$ results. Most important, 
even in this case the preferred $n_z$ values point to a redshift 
evolution $n_z>0$. 

More generally, our framework relies on several simplifying choices: 
Our MCMC analysis relies on discrete prior grids, which limits the 
precision with which parameter degeneracies can be mapped. We have 
verified that the adopted step sizes yield valid results for the 
current sample size, but finer grids will be warranted as statistical 
uncertainties shrink with larger FRB samples. Furthermore, we have explored 
degeneracies with $\mu_{\rm host}$, $H_0$, and $F$, but additional 
parameters -- such as the slope of the FRB luminosity function or 
the modeling of the Milky Way DM contribution -- may introduce 
further correlations that we have not accounted for here 
\citep{James2022,Hoffmann2025}. These are expected to be subdominant 
given the current sample size, but will become relevant as 
measurement precision improves. Finally, we note that ${\rm DM_{host}}$
encodes the \textit{joint} evolution of host gas density, path
length, and FRB-host geometry; separating these contributions is not
possible with current data alone, and will require independent
constraints on host size and morphology evolution to fully interpret
$n_z$ in terms of ionized gas density evolution \citep[see also][]{Orr2024}.

\subsection{Comparison to Previous Work} 

As mentioned in Section~\ref{sec:intro}, \cite{Sang2025} recently  
analyzed 117 localized host FRB, including our DSA and CRAFT samples, 
and with the addition of other MeerTRAP, ASKAP and CHIME \cite{Amiri2025} 
sources. These authors found a correlation (Pearson coefficient $r=0.52$) 
between redshift and host DM equivalent to  $n_z\approx0.9$ in our formalism, 
which  closely matches our derived value for DSA. Given that our CRAFT and joint 
results may be impacted by the single $z>1$ CRAFT FRB with notably high DM, 
this value appears reasonable and well within our large 1\,$\sigma$ uncertainty. 
\cite{Sang2025} calculated host DM values by subtracting the Milky Way and a 
model of the cosmic component to the total FRB DM, with the largest 
uncertainties arising from the modeling of the cosmic term. Future work is 
required to confirm or refute this value. 

The likelihood analysis by \cite{Kumar2025}  on 65 FRBs 
largely overlapping with our sample found $n_z=0.24\pm 1.92$, but did not 
account for instrumental (observational)  effects. The 
potential for biases when not taking these effects into account  
was emphasized in a similar analysis on $H_0$ by \cite{James2022}, 
and our work has also stressed the strong dependence of the results on 
the shape of the underlying 
$p({\rm DM_{EG}},z)$ distribution.

\cite{Bernales2025} presented a comprehensive analysis of DM$_{\rm host}$ 
measurements based on H$\alpha$ emission for one MeerTRAP and 11 CRAFT 
FRB and they found $n_z=0.3\pm1.7$. Importantly, they report positive 
correlations of the host DM -- from the ISM, as well as from the halo --  
with stellar mass and star-formation rate. However, they argued that this is 
expected since most of their hosts lay on the galaxy star-formation main sequence 
and, therefore, stellar and halo mass, star formation, emission and 
emission-based host DM, all correlate. Future work with a large host sample  
will reveal whether this correlation exists for the whole host population.

\section{Conclusions}\label{sec:conclusions}

We have explored the use of fast radio bursts (FRBs) as a novel probe of the 
redshift evolution of ionized gas in the whole galaxy-halo system, 
through the redshift dependence of the FRB host dispersion measure, 
${\rm DM_{host}}(z) \propto (1+z)^{n_z}$. The host DM 
describes the total electron column density across all 
    ionized phases along the FRB 
    host sightline, providing a unified tracer of the ionized gas evolution that 
    complements phase-specific probes such as  emission lines, 
    quasar absorption-line spectroscopy, and X-ray/SZ observations. 
    Unlike these conventional diagnostics, FRBs require no
assumptions about photoionization state, line selection, or
availability of background sources, and are equally sensitive to all
ionized phases from the ISM to the CGM.
Using a forward-modeling framework that accounts for instrumental 
 effects, and applying it to 90 localized FRBs (69 with 
confirmed host redshifts) from the DSA and ASKAP/CRAFT ICS surveys, 
our main findings are:

\begin{itemize}

    \item \textbf{Evidence for a host DM redshift evolution.} 
    Our joint inference on DSA and CRAFT FRBs yields 
    $n_z = 1.62^{+1.48}_{-1.57}$, ruling out the non-evolving case 
    ($n_z = 0$) at $1\,\sigma$. Both individual datasets independently also 
    favor the existence of an evolution $n_z>0$, but their large uncertainties 
    cover the value $n_z=0$. This is the first
measurement of $n_z$ accounting for instrumental selection effects, 
and represents a proof of concept that the method is viable with
current data.

    \item \textbf{Implications for FRB science.} An evolving host DM impacts various FRB-based calculations: 
     (i) it generally overestimates  
    DM$_{\rm EG}$-based redshifts  
    if ignored, by up to $\Delta z \approx 0.3$ for
    DM$_{\rm EG}\sim1000-2000\,{\rm pc\,cm^{-3}}$; (ii) it propagates into constraints on cosmological 
    parameters (e.g., $H_0$, baryon density distribution, helium reionization) and 
    astrophysical parameters (e.g., galaxy feedback) through the cosmic DM; 
    and (iii) combined with host galaxy type and properties, it may aid 
    in constraining FRB progenitor channels. We recommend that future FRB inference analyses take a potential evolution into account, e.g., 
    marginalizing over $n_z$ or adopting our result as an
    informative prior.

    \item \textbf{Key parameter degeneracy: $n_z$ and $\mu_{\rm host}$.} 
    The evolution index $n_z$ is most degenerate with the mean host DM value, 
    $\mu_{\rm host}$, particularly for samples containing high-redshift FRBs. 
    This degeneracy must be accounted for to obtain unbiased results; $n_z>0$ 
    drives $\mu_{\rm host}$ toward lower values, as 
    illustrated by our result $\log\mu_{\rm host} = 2.03$, compared to 
    $\log\mu_{\rm host} = 2.18$ from analyses assuming no evolution, a $\sim30\%$ shift in the mean host DM that
    propagates directly into the cosmic DM budget and all parameters
    derived from it, although 
    both are consistent within $1\,\sigma$.

    \item \textbf{High-redshift FRBs are the strongest discriminants.} 
    Differences between evolution scenarios grow with redshift, making 
    high-$z$ FRBs the most sensitive indicators of $n_z$. MeerTRAP in 
    coherent mode, with its higher sensitivity and larger  fraction of 
    expected high-redshift detections, is the best suited current instrument 
    for precisely localizing these FRBs.
    About 100 (300/350) MeerTRAP in coherent mode 
    (DSA/CRAFT) FRB hosts with measured redshift are required to reduce the $n_z$
    uncertainties down to $\sim0.7$.

\item \textbf{Physical interpretation in the context of galaxy evolution.} A well-constrained $n_z$ will 
    enable  comparisons with the density evolution of individual 
    ionized phases and with galaxy scaling relations, revealing the dominant 
    ionized gas phase and the 
    driver of the overall  evolution in the galaxy-halo system. This 
    will inform models of how ionized gas in galaxies and their surroundings  
    co-evolve across cosmic time. However,  ${\rm DM_{host}}$ encodes the joint evolution of gas
    density, path length, and FRB-host geometry -- contributions that
    cannot currently be separated. As an illustration only,
     assuming a star-forming host with 
    size evolution $r \propto (1+z)^{-1}$ and $n_z = 1.62$ implies an 
    electron density evolution $n_e \propto (1+z)^{2.6}$, consistent with 
    the cosmic star formation rate evolution and potentially suggestive of 
    the diffuse ionized gas (DIG) in the ISM  as 
    the dominant ionized phase. Robust conclusions of this kind await 
    tighter constraints on $n_z$ from future, larger FRB samples, combined with independent
    constraints on host size and morphology evolution (Khrykin et al., in prep.).

\end{itemize}

This work establishes a framework for measuring host DM evolution from 
observational data and highlights the importance of including $n_z$ 
as a free parameter in FRB inference analyses. Future work will apply 
this framework to larger samples -- enabled by upcoming instruments 
(CHIME/FRB Outriggers, CHORD, DSA-2000) and spectroscopic surveys (DESI, EUCLID) -- and will explore correlations between the DM 
evolution and host galaxy type (e.g., star forming or quenched) and physical properties, further 
unlocking the potential of FRBs as novel probes of galaxy evolution across 
cosmic time.

\begin{acknowledgments}
We are grateful to Clancy W. James for discussions, and his assistance with the \texttt{zDM} software package that made this analysis  
possible. We also thank  the colleagues of the Fast and Fortunate for FRB 
Follow-up (F4) and MeerTRAP teams for comments and suggestions. 
The connection between an undetected FRB population at low redshift and high-redshift FRBs with host DM excess was first suggested by a referee of \cite{Caleb2026}, who we thank for such an idea.
\end{acknowledgments}

\appendix

\section{Degeneracies between \MakeLowercase{n$_\MakeLowercase{z}$} and  $\sigma_{\rm \MakeLowercase{host}}$ }\label{sec:applike}

\begin{figure*}\centering
\includegraphics[width=0.33\textwidth]{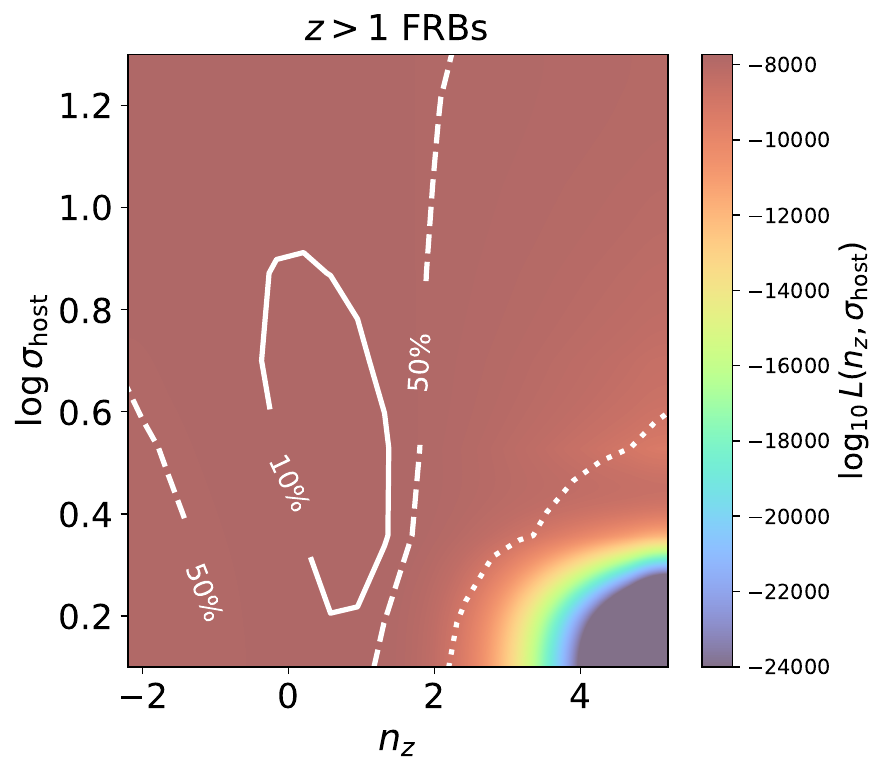}\includegraphics[width=0.33\textwidth]{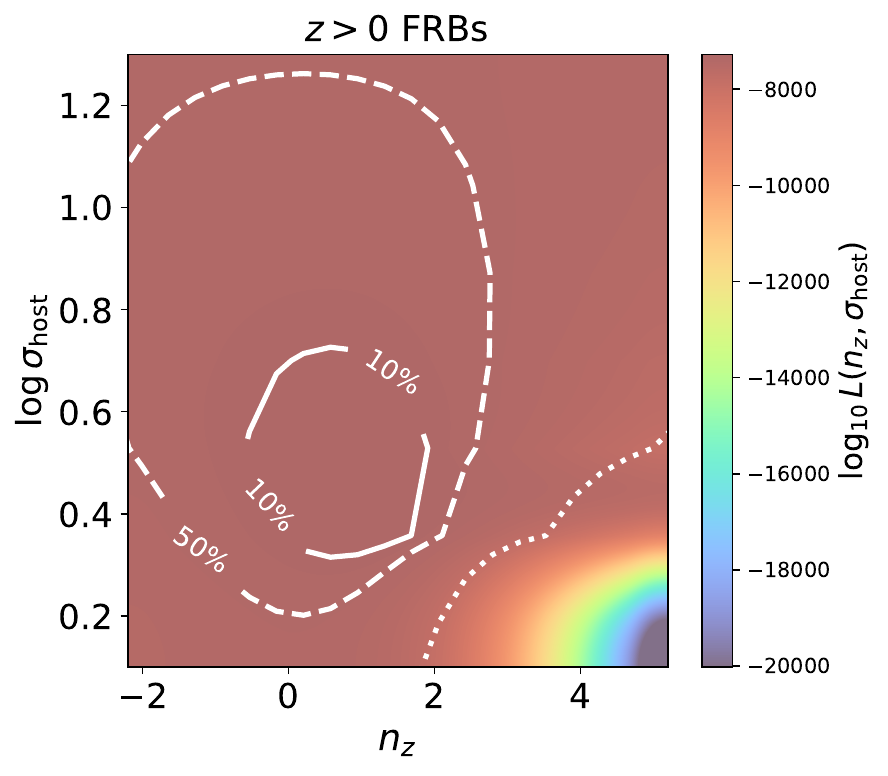}\includegraphics[width=0.33\textwidth]{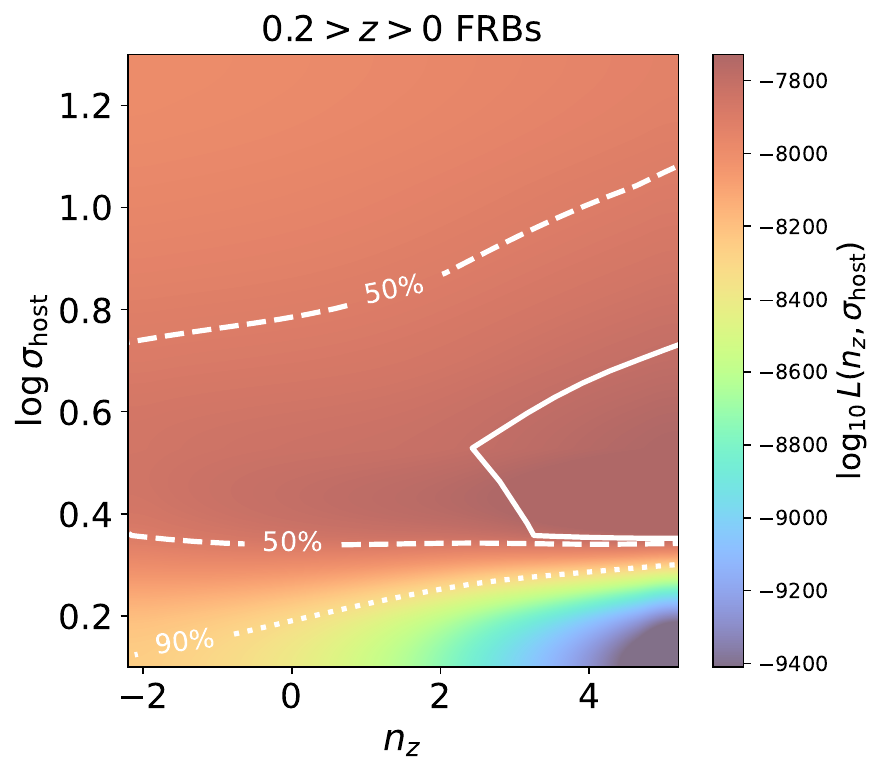}
\includegraphics[width=0.33\textwidth]{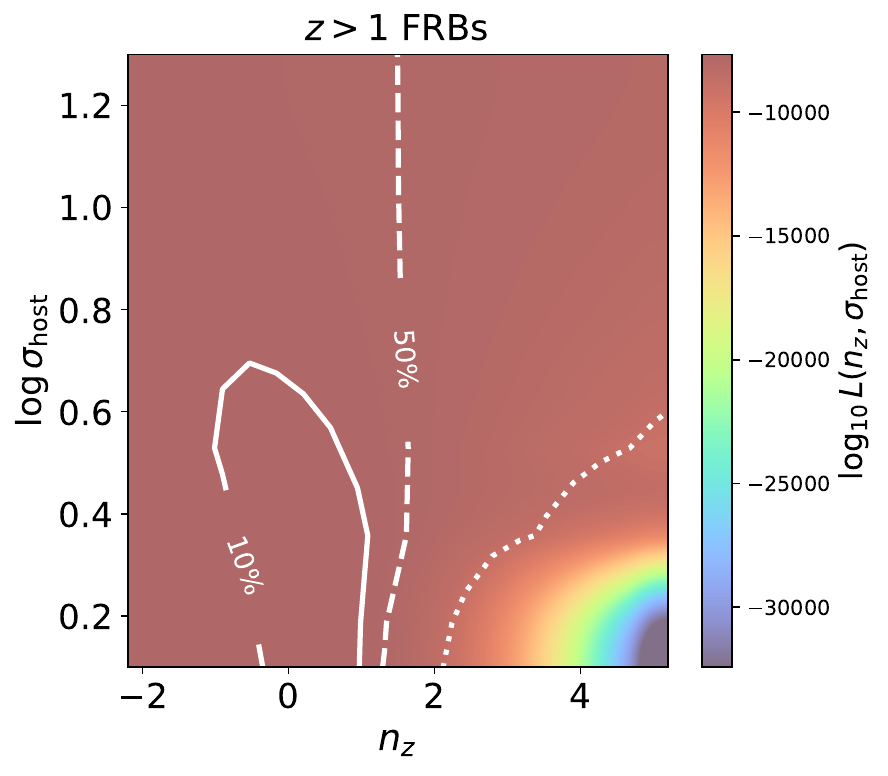}\includegraphics[width=0.33\textwidth]{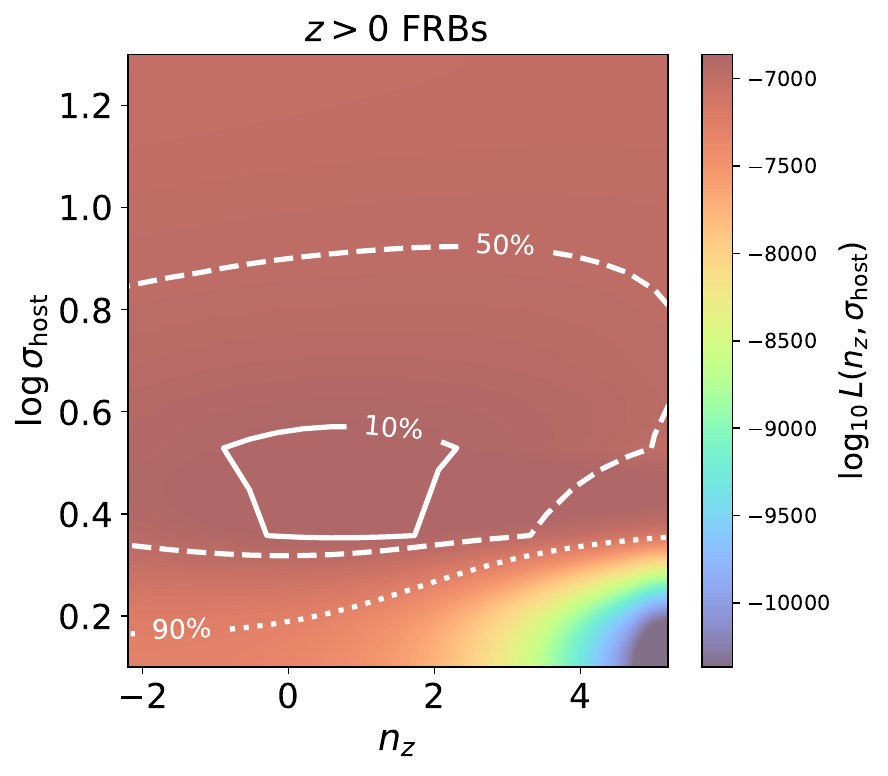}\includegraphics[width=0.33\textwidth]{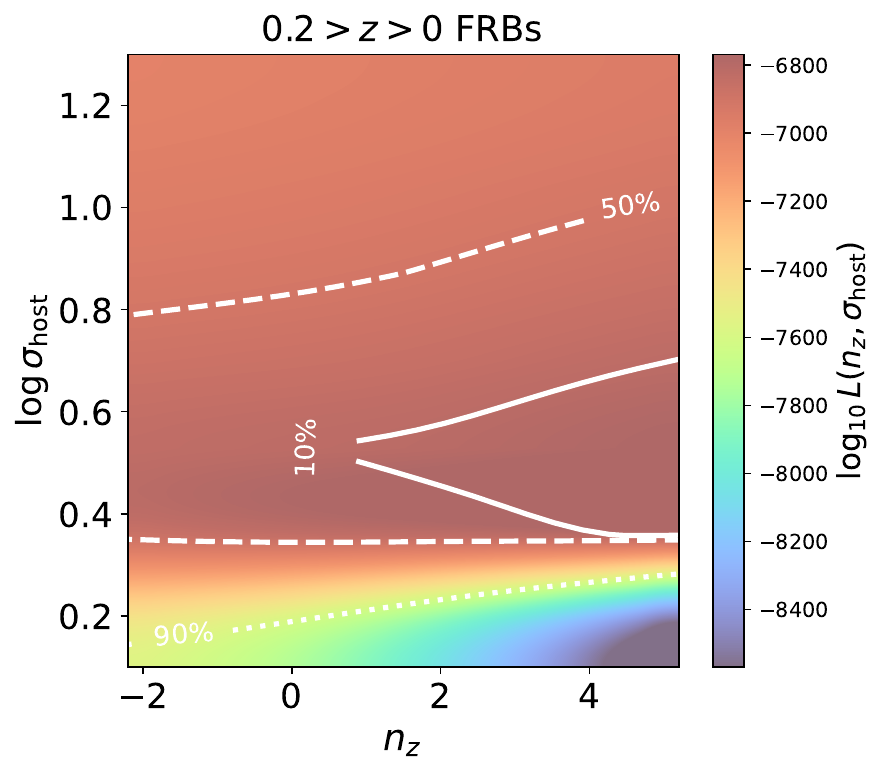}
\includegraphics[width=0.33\textwidth]{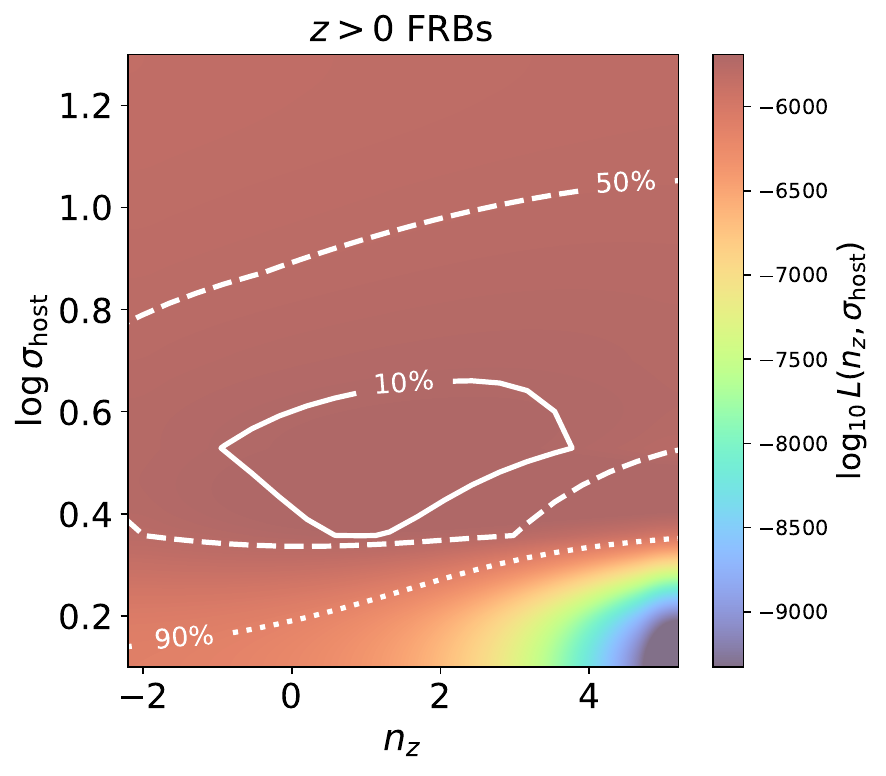}\includegraphics[width=0.33\textwidth]{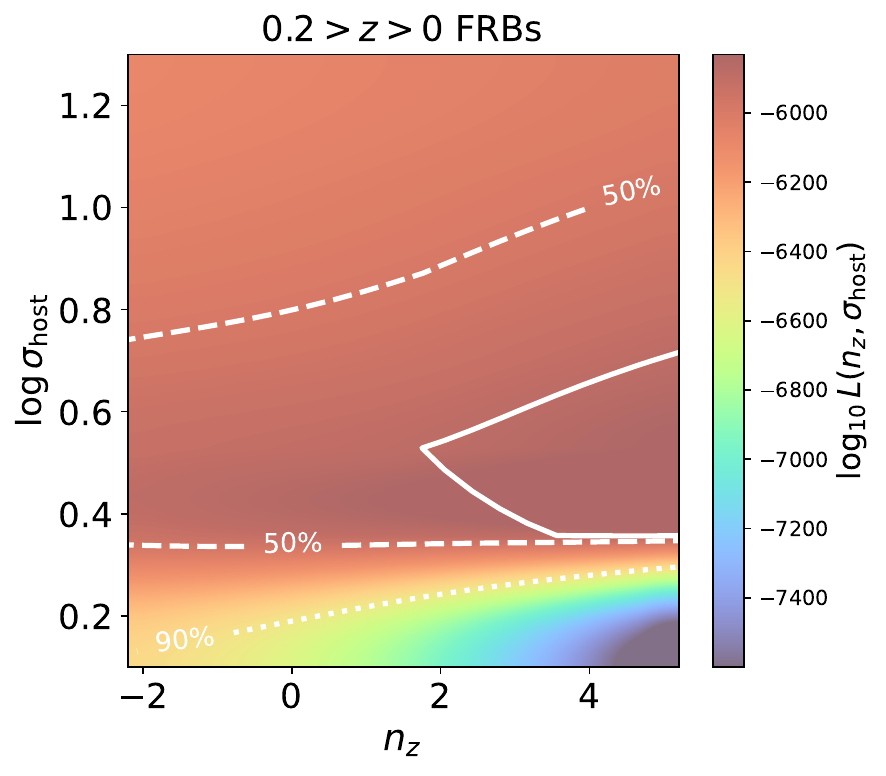}
\caption{Same as Figure~\ref{fig:degen} but replacing $\mu_{\rm host}$ by 
$\sigma_{\rm host}$. No significant correlations between $\sigma_{\rm host}$ and $n_z$ appear visible.}
\label{fig:sig}
\end{figure*}

Results from the likelihood calculations for the MeerTRAP, DSA and ASKAP/CRAFT ICS 
experiments with $n_z=1$ are presented in Figure~\ref{fig:sig}.

\clearpage
\bibliography{sample7}{}
\bibliographystyle{aasjournalv7}

\end{document}